\documentclass[12pt]{article}

\RequirePackage{amsthm,amsmath,amsfonts,amssymb}
\RequirePackage[numbers,sort&compress]{natbib}
\RequirePackage[colorlinks,citecolor=blue,urlcolor=blue]{hyperref}
\RequirePackage{graphicx}

\usepackage{enumerate}
\usepackage{authblk}

\usepackage{url}            
\usepackage{orcidlink}

\newcommand{\blind}{0}

\usepackage[top=.8in,left=.9in,right=.9in,bottom=.9in]{geometry}

\usepackage{booktabs}       
\usepackage{nicefrac}       
\usepackage{microtype}      
\usepackage{dsfont}
\usepackage{algorithm}
\usepackage{algpseudocode}
\usepackage{tikz,xcolor}
\usepackage{comment}
\usepackage{chngcntr}
\usepackage{euscript,mathrsfs}
\usepackage{caption}

\newtheorem{theorem}{Theorem}[section]
\newtheorem{lemma}[theorem]{Lemma}
\newtheorem{corollary}[theorem]{Corollary}
\newtheorem{prop}[theorem]{Proposition}

\newtheorem{definition}[theorem]{Definition}

\usetikzlibrary{arrows.meta}

\definecolor{darkgreen}{rgb}{0.0, 0.3, 0.0}
\definecolor{darkred}{rgb}{ 0.6,0, 0.0}
\definecolor{darkblue}{rgb}{0,0,0.5}

\newcommand{\indep}{\perp \!\!\! \perp}

\renewcommand{\S}{\EuScript{S}}
\newcommand{\wt}[1]{\widetilde{#1}}
\newcommand{\p}[1]{\mathbb{P}\left(#1\right)}

\newcommand{\E}{\mathbb{E}}
\newcommand{\R}{\mathbb{R}}
\renewcommand{\P}{\mathbb{P}}
\newcommand{\sgn}{\operatorname{sgn}}

\newcommand{\U}{\mathcal{U}}
\newcommand{\Y}{\mathcal{Y}}
\newcommand{\V}{\mathcal{V}}

\newcommand{\Hg}{H_0}

\newcommand{\eu}[1]{\EuScript{#1}}
\newcommand{\diff}{\mathrm{d}}
\newcommand{\F}{\mathcal{F}}
\newcommand{\f}{\mathfrak{f}}
\newcommand{\ee}{\mathrm{e}}

\begin{document}

\def\spacingset#1{\renewcommand{\baselinestretch}%
	{#1}\small\normalsize} \spacingset{1}


\if0\blind
{
\title{\bf Stable Distillation and High-Dimensional Hypothesis Testing}

\author[1,2]{\orcidlink{0000-0002-2049-3389}Ryan Christ\thanks{Corresponding author: Ryan Christ, ryan.christ@yale.edu}}
\author[1,2]{\orcidlink{0000-0003-4442-6655}Ira Hall\thanks{The first and second authors were supported in part by NIH Grant UM1 HG008853/HG/NHGRI \newline NIH HHS/United States.}}
\author[3]{\orcidlink{0000-0003-3044-5433}David Steinsaltz\thanks{The third author was supported in part by BBSRC grant number BB/S001824/1.
        For the purpose of Open Access, the authors have applied a CC BY public copyright license to any Author Accepted Manuscript version arising from this submission.}}
\affil[1]{Center for Genomic Health, Yale School of Medicine, New Haven, CT, USA}
\affil[2]{Department of Genetics, Yale School of Medicine, New Haven, CT, USA}
\affil[3]{Department of Statistics, University of Oxford, Oxford, UK}

%
%

	\maketitle
} \fi

\if1\blind
{
	\bigskip
	\bigskip
	\bigskip
	\begin{center}
		{\LARGE\bf Stable Distillation and High-Dimensional Hypothesis Testing}
	\end{center}
	\medskip
} \fi

\bigskip
\begin{abstract}
			While powerful methods have been developed for high-dimensional hypothesis testing assuming orthogonal parameters, current approaches struggle to generalize to the more common non-orthogonal case. 
			We propose \emph{Stable Distillation} (SD), a simple paradigm for iteratively extracting independent pieces of information from observed data, assuming a parametric model. 
			When applied to hypothesis testing for large regression models, SD orthogonalizes the effect estimates of non-orthogonal predictors by judiciously introducing noise into the observed outcomes vector, yielding mutually independent p-values across predictors. 
			Generic regression and gene-testing simulations show that SD yields a scalable approach for non-orthogonal designs that exceeds or matches the power of existing methods against sparse alternatives. While we only present explicit SD algorithms for hypothesis testing in ordinary least squares and logistic regression, we provide general guidance for deriving and improving the power of SD procedures.
\end{abstract}

\noindent%
\vfill

\newpage
\spacingset{1.9} 

	\hypertarget{introduction}{%
		\section{Introduction}\label{introduction}}
	\subsection{The problem of global null hypothesis testing}

	Consider the problem of testing a null hypothesis $\Hg : \left\{\theta = \theta_0 \right\}$ about a $p$-dimensional parameter $\theta$ given some observed data $Y$ drawn from a known family of distributions $F_\theta$.  
	$\Hg$ may be motivated by a scientific hypothesis, such as whether any measured environmental toxins modulate disease risk \cite{su2016association}, or arise during model selection, where we want to test whether we should add any more features to a given model \cite{lockhart2014significance}. 
	
	Here we focus on procedures with power against sparse alternatives: in modern applications ranging from genetics and communications to signal detection and compressive sensing, $p$ may be in the millions, while only very few elements of $\theta$ actually depart from $\theta_0$ under the alternative hypothesis \cite{ACCP11}. 
	Throughout we consider the general case of non-orthogonal parameters: in most applications the likelihood implied by a given $F_\theta$ does not factorize according to $\theta$. 
	In the regression context, this refers to the fact that coefficient estimates are dependent unless there is orthogonality among their corresponding predictors. 
	We also allow for nuisance parameters: often $F$ is indexed by additional parameters alongside $\theta$ that must also be estimated from $Y$. 
	
	We propose a stochastic decoupling framework we call \emph{Stable Distillation} (SD).
	Informally, SD provides a principled paradigm for repeatedly and interactively extracting information from $Y$, sometimes called ``double dipping''. 
	SD extracts some information $U$ (relevant to $\Hg$) from $Y$ while injecting noise into $Y$ in a way that maintains the distribution of $Y$ under $\Hg$ and forces the updated $Y$ to be independent of $U$. Iterating this procedure enables the accumulation of independent pieces of information about $\Hg$ across iterations, yielding powerful hypothesis tests, especially in the aforementioned sparse-signal context. Although the framework we present is much more general, given our focus in this paper on regression applications, it will often be helpful to think about each extracted $U$ as a p-value relevant for a particular parameter $\theta_j$.

	From an intuitive perspective, SD ``hides'' the data $Y$ from the statistician or downstream statistical methods. However, the statistician is allowed to interactively make queries of --- extract information from --- the data. If we can assume an adequate model $Y\sim F_\theta$, then SD guarantees that the response to each query will be independent. Thus, if the queries are carefully posed, the statistician may obtain new, helpful pieces of information with each query.
	
	SD is distinct in both essence and effect from existing work. While the idea of ``hiding'' the data $Y$ has become quite popular within the interactive hypothesis testing literature \cite{lei2018adapt, fithian2020, duan2020familywise, duan2020interactive}, SD posits a likelihood-specific procedure that explicitly and sequentially updates $Y$ itself. While this means that more work is required to derive a valid SD procedure for a  given likelihood, as evidenced by exciting recent work on data fission \cite{leiner2023data} and stable algorithms \cite{ZJ23}, likelihood-specific procedures can yield power advantages that are not accessible by more generic methods.
	
	While data fission and stable algorithms may appear superficially similar to SD, they both use randomization to produce independence between \emph{stages} of inference rather than between model parameters. 
	To understand this distinction, note that data fission and stable algorithms are both in their own way strengthened versions of Cox's classic data splitting procedure for large regression models \cite{leiner2023data, ZJ23}: some samples are used to select potentially interesting variables, and then other samples are used to estimate the effects of those selected predictors \cite{cox_1975}. 
	
	SD has no such analogy to data splitting. Rather than performing a one-shot bi-partitioning of the information in $Y$ into a piece for parameter selection and piece for parameter inference, in a regression context, SD partitions the information into $p$ independent pieces. Effectively, all available information and samples are used when SD extracts evidence about each $\theta_j$. Using noise to slice the inference procedure into $p$ pieces along $\theta$ is not costless, of course, but it presents a usefully novel portfolio of risks and benefits. In particular, it allows SD to leverage outlier tests to test the global null in the sparse-signal setting --- an approach that is not open to existing methods. It also allows SD to incorporate a new class of filters, introduced via a simple example in Algorithm \ref{alg:simple_quantile_filter} and fully described in Section \ref{sec:generalfiltering}, to increase power. These filters are distinct from those used in interactive hypothesis testing \cite{lei2018adapt,duan2020interactive}. 
	
	The novel design of SD is also reflected in the general coupling framework presented in Section \ref{sec:generalEFR}, which generalizes the type of factorization at the core of data fission and allows us to develop an SD procedure for logistic regression based on coupled random walks in Section \ref{sec:LR_example}. We further situate SD within the relevant literature in Section \ref{sec:literature}, after we have explained its essential features. Applications of SD to more complex inference problems beyond hypothesis testing, such as variable selection, are left to future work. 
	
	The abstract definition of SD given in Section \ref{sec:sd_def} does not immediately lead to useful procedures for particular classes of null and alternative hypotheses. We focus on providing explicit procedures for the common case of comparing nested regression models, where $\theta$ corresponds to some subset of coefficients. In Section \ref{sec:motivation} below, we motivate SD in the context of the sizable literature on high-dimensional hypothesis testing for ordinary least squares regression. We present explicit SD procedures for ordinary least squares regression in Section \ref{sec:ols_example} and logistic regression in Section \ref{sec:LR_example}. In the context of regression, applying SD yields an independent p-value $U$ for each predictor, allowing direct application of an outlier test such as Tukey's higher criticism to test $\Hg$ with power against sparse alternatives \cite{tukey_hc,donoho2004higher}. The simulations in Section \ref{sec:power_sims} demonstrate that this approach yields considerable power gains in settings with sparse alternatives and high correlation among predictors. Genomics simulations in Section \ref{sec:experiments} show that these gains carry over to the context of gene testing.
	We provide general guidance for obtaining explicit and powerful SD procedures for testing $\Hg$ under any $F_\theta$ throughout.

	\subsection{Notation}  \label{sec:notation}
	Throughout the paper we will be defining distributions within a measurable space $(\Omega,\mathcal{F})$.
	Unless otherwise indicated, the probability operator $\P$ and the expectation operator $\E$ will be assumed to refer to the null distribution defined by what we are calling, in the relevant context, the null hypothesis $\Hg$.
	Denote the cdf of the Beta distribution with parameters $a,b$ (with mean $a/(a+b)$) by $F_{a,b}(x)$. For a matrix $M \in \mathbb{R}^{n \times m}$ where $n \geq m$, $P_M \in \mathbb{R}^{n \times n}$ will denote a projection matrix onto the subspace spanned by the columns of $M$; $P^\bot_M= I - P_M$, a projection matrix onto the orthogonal subspace. 
	For a set of columns $B$ we let $M_{\cdot B}$ denote the matrix $M$ subsetted down to the corresponding columns; for a set of rows $B$ we let $M_{B\cdot}$ denote, the corresponding rows. For a positive integer $n$ we will abbreviate the set $\left\{1,\ldots,n\right\}$ as $[n]$, and use $\mathcal{S}_n$ to denote the set of permutations on $[n]$.

	\subsection{Sparse-signal OLS models}\label{sec:motivation}
	
	For an observed response $Y \in \mathbb{R}^n$, a background covariate matrix $A \in \mathbb{R}^{n \times q}$ where $q < n$, and a design matrix $X \in \mathbb{R}^{n \times p}$, the OLS model is
	\begin{equation} \label{model}
		Y = A\alpha + X \beta + \sigma \epsilon
	\end{equation}
	where $\epsilon \overset{\mathrm{iid}}{\sim} N(0,1)$ for all $i = 1,\ldots,n$ and $\sigma$ is an unknown scalar. In this regression context, we will consider the null hypothesis $\Hg : \left\{\beta_1 = \beta_2 = \ldots = \beta_p = 0\right\}$. Here, $p$ may be very large: possibly greater than $n$ and perhaps in the millions. 
    Throughout, we will assume that $P^\bot_A X_{\cdot j} \neq 0$ for all $j = 1,\ldots,p$, but make no further assumptions about the structure of the $X_{\cdot j}$ or the identifiability of each $\beta_j$. The predictors encoded in the columns of $X$ may be strongly (even perfectly) correlated, making the resulting $\hat{\beta}_j$ dependent. Also note the presence of nuisance parameters, $\alpha$ and $\sigma$. 
	
	We are particularly interested in the ``sparse-signal'' setting, where the number of active predictors, $a = \left|\left\{X_{\cdot j} : \beta_j \neq 0 \right\}\right|$, is very small compared to $p$. Note, our SD approach remains valid for any $\beta$. Hypothesis testing in this sparse-signal, high-dimensional context is common in modern applications. Sometimes predictors are collected with an eye toward a specific hypothesis: early pandemic detection, signal processing, and clinical trial rescue \cite{donoho2004higher}. 
	In many other scientific studies, predictors are collected without any known relevance: screening drug compounds for activity against some disease process \cite{michelini2010cell}. Testing $\Hg$ is also an essential tool in model selection \cite{ACCP11}. For example, a financier may test whether there are any asset prices or economic variables from a large class that might be worth including in their forecast model. The association signals may be sparse and weak, and the asset prices may be strongly dependent, but they might be worth exploring in order to achieve a small increase in returns.
	
	For the OLS model, the classic likelihood ratio test against $\Hg$ (ANOVA) and related ``sum-of-squares-type'' statistics naturally account for non-orthogonality among parameters and the presence of nuisance parameters \cite{goeman2006testing,chen2019two,guo2016tests}. While these statistics are well powered against ``dense" alternatives, where $a$ is not so small compared to $p$, these statistics have little to no power against sparse alternatives \cite{ACCP11,he2021asymptotically}. 
    If $X$ followed an orthogonal design, and there were no nuisance parameters, then the standard method to test $\Hg$ against the sparse alternatives would be to first regress $Y$ on each $X_{\cdot j}$ separately to obtain an independent p-value for each $X_{\cdot j}$ and then test $\Hg$ with a goodness-of-fit test such as higher criticism that targets sparse signals \cite{donoho2004higher}.
	Compared to ``minimum"-based approaches, such as testing the minimum p-value with Bonferroni correction or Holm's method,
	higher criticism and related tests in the General Goodness of Fit test family have more power when, roughly speaking, there are a handful of active predictors with modest effects \cite{donoho2004higher, hz19, zhang2022general}. 
	
	Unfortunately, higher criticism and related outlier tests do not have known null distributions under non-orthogonality and the presence of nuisance parameters. Directly simulating the null distribution via Monte Carlo, as done in \cite{ACCP11}, is prohibitive when $p$ is large and intractable in screening applications that test many $\Hg$. Such screening applications require precise estimates of the tail of the null distribution to account for multiple testing. For example, a standard challenge in genomics is to screen approximately 20,000 human genes to find those that might influence a disease of interest with a Bonferroni threshold around $2.5 \times 10^{-6}$.
	Here, association between the disease and the gene is typically tested by comparing nested regression models where the larger model includes all of the genetic variants within a given gene. Modern sequencing datasets may easily have tens of thousands of variants in a single gene, and the sample size $n$ these datasets is now reaching into the millions \cite{allofus}. We apply SD to gene testing in Section \ref{sec:experiments}. 
	
	As an alternative to higher criticism, impressive recent work on higher-order-norm U-statistics, \cite{he2021asymptotically} has sought to bridge the gap between statistics that are powerful against dense and sparse alternatives by considering the joint distribution of $||\hat{\beta}||_q$ across several positive integer $q$ and $q = \infty$. Unfortunately, this work does not admit nuisance parameters and only an asymptotic null distribution is available, which makes calculating p-values at the scale $10^{-6}$ unreliable for applications like gene testing. Moreover, convergence to this null distribution requires difficult-to-verify limits on the correlation among the predictors that are violated in practice: in gene testing applications $||\hat{\beta}||_2$ is observed to converge to a generalized chi-square distribution rather than a Gaussian \cite{lumley2018fastskat}.
	
	Various attempts have been made to extend higher criticism to non-orthogonal predictors while avoiding Monte Carlo simulation. The direct approach --- rotating the Gaussian test statistics to produce new orthogonal test statistics, hence yielding independent p-values \cite{IHC} --- destroys any sparsity in the original design, causing the test to lose power rapidly as the correlations increase \cite{GHC}. Generalized Higher Criticism \cite{GHC} takes a different, in principle more powerful, approach, deploying ingenious computations to adjust the null distribution based on the correlation structure of dependent test statistics \cite{GHC}. Unfortunately, this method scales like $\mathcal{O}(p^3)$, making its use in problems where $p$ is much larger than $100$ impractical.
	
	There are several scalable off-the-shelf approaches to combining dependent p-values, such as the classic Simes procedure, the Cauchy Combination Test, and recent averaging-based p-value methods \cite{simes1986improved,hochberg1988sharper,vovk2020combining,liu2020cauchy}. The Cauchy Combination Test has proven to be relatively powerful and is widely used in genomics applications \cite{liu2020cauchy}.
	However, these procedures pay a substantial price in power to maintain type-1 error control in the presence of dependent p-values. Generalized Higher Criticism and related members of the General Goodness of Fit test family suffer the same problem.
	For intuition, consider the classic multiple testing problem of regressing some outcome vector onto each predictor, one-at-a-time to achieve some set of p-values. Assume two of these observed p-values are particularly small. If they correspond to nearly identical predictors, then the evidence against $\Hg$ given by these each p-value is redundant: there is really only one outlier, not two. However if these two small p-values correspond to orthogonal predictors, they provide independent evidence against $\Hg$ and their signals may be combined to boost power. 
	Off-the-shelf approaches allowing for dependent p-values cannot distinguish between these two cases. In other words, they cannot distinguish observations where all of the outlying test statistics come from a single tightly correlated cluster of predictors from observations where the outlying test statistics correspond to orthogonal predictors.
	In contrast, our SD procedures make this distinction in the process of orthogonalizing the $\hat{\beta}_j$. It is this independence among p-values that ultimately allows higher criticism, or any outlier test used after distillation, to out-perform generic methods that allow for dependent p-values. As we show in Section \ref{sec:power_sims}, this allows SD to achieve significantly more power then the Cauchy Combination Test when there are clusters of strongly correlated predictors.

	\section{Stable Distillation} \label{sec:sd_def}
	All of the random variables we describe in this section will be assumed to take values in a Polish space, with the Borel $\sigma$-algebra, but we will generally not need to specify these spaces further.
	Given observed data $Y$ drawn from some family of distributions $F_\theta$ parameterized by $\theta$, a SD process is a stochastic process $(Y^{(l)},U^{(l)})_{l=1}^L$ initialized with the observed data $Y^{(1)} = Y$. 
	In a regression context, the time steps $l$ may correspond to predictors, which is to say, to columns of the design matrix $X$.
    
    	\begin{figure}[t]
		\centering
		\begin{tikzpicture}[node distance=.5cm,scale=1.5]
			\input{hmarkov_diagram.tex}
		\end{tikzpicture}
		\caption{\textbf{Bayesian network structure for an HMM.} Arrows represent independence of a node from all of its nondescendants, conditioned on its parents, as in Definition 3.1 of \cite{KF09}.} \label{F:HMM} 
	\end{figure}
    
	An SD process follows the conditional independence structure of an abstract hidden Markov model (HMM), sketched in Figure \ref{F:HMM}, where the data play the role of ``hidden states'', and test statistics from that data are the ``observed states''. In line with Figure \ref{F:HMM}, here we define an HMM as a stochastic process $(Y^{(l)},U^{(l)})_{l=1}^L$ such that $Y^{(l)}$ is independent of $\big((Y^{(l)})_{l=1}^{l-2},(U^{(l)})_{l=1}^{l-1}\big)$ given $Y^{(l-1)}$ and $U^{(l)}$ is independent of $\big((Y^{(l)})_{l=1}^{l-1},(Y^{(l)})_{l=l+1}^{L},(U^{(l)})_{l=1}^{l-1},(U^{(l)})_{l=1+1}^{L}\big)$ given  $Y^{(l)}$. However, many of the results and algorithms used for HMMs do not apply to SD processes since SD processes must also satisfy a crucial marginal independence condition: each $U^{(l)}$ must be independent of $ Y^{(l+1)}$.
	\begin{definition} \label{def:sd}
		Given an initial $Y^{(1)} = Y \sim F_{\theta_0}$, a {\em stable distillation} (SD) process is a hidden Markov model $(Y^{(l)},U^{(l)})_{l=1}^L$ where $Y^{(l)} \sim F_{\theta_0}$ for all $l \in \left\{2,\ldots,L\right\}$ and $U^{(l)} \ \indep  \ Y^{(l+1)}$ for all $l \in \left\{1,\ldots,L-1\right\}$.
	\end{definition}
    
	The HMM structure of SD may be understood intuitively: the observed data $Y^{(1)}$, and modified copies of that data, $Y^{(2)},\ldots, Y^{(L)}$, are ``hidden'' from the researcher. 
	The researcher, and any downstream statistical procedures, only get to observe statistics $\left(U^{(l)}\right)_{l=1}^L$ ``emitted'' by the process. Typically, each $U^{(\ell)}$ will be a p-value or set of p-values. This explicit masking of the data preserves information, so that later inference steps remain unbiased. 
	While Definition \ref{def:sd} requires that the marginal distribution of each $Y^{(l)}$ be stable, it does not assume that the Markov chain $\left(Y^{(l)}\right)_{l=1}^L$ is time-homogeneous. Critically, Definition \ref{def:sd} guarantees independence among the $(U^{(l)})_{l=1}^L$.
	\begin{lemma}
		\label{L:Uindependence}
		If $(Y^{(l)},U^{(l)})_{l=1}^L$ is an SD process, then $ U^{(1)}, U^{(2)} \ldots, U^{(L)}$ are mutually independent.
	\end{lemma}
	\noindent This makes it relatively straightforward to obtain a calibrated inference procedure that accumulates information across $(U^{(l)})_{l=1}^L$. 

    At the start of each iteration $l$, one may use the emitted statistics $U^{(k)}$ from in prior iterations $(k<l)$ in selecting which information to extract from $Y^{(l)}$. In the context of the OLS model where each $U^{(l)}$ may correspond to a p-value for a given predictor, one may use the previous $U^{(k)}$ to select the next predictor to distill. While this ``human in the loop'' feature provides a basis for interactive hypothesis testing, which has been a major recent focus in the literature \cite{lei2017star,duan2020familywise}, the examples we present in this paper do not make use of this freedom.     
    
    \section{EFR-SD via Factorization}\label{sec:sd}

    The challenge lies in constructing explicit SD processes that meet the requirements of Definition \ref{def:sd} for a given likelihood $F_\theta$ and maintain power for inference. We propose a useful, albeit more constrained, framework, Extraction-Filtration-Reconstitution Stable Distillation (EFR-SD), for obtaining explicit SD procedures. Intuitively, the approach involves judiciously injecting noise into $Y^{(l)}$ to obtain the next $Y^{(l+1)}$. This noise ensures stability $\left(Y^{(l+1)} \sim F_{\theta_0} \right)$ and serves as ``payment'' for observing $U^{(l)}$. Here we consider the special case where we have an explicit factorization available via a function $\psi$. We present the more general framework, EFR-SD via coupling, in Section \ref{sec:generalEFR}.
    
    \subsection{The Factorization Framework} \label{sec:independentEFR}
\begin{figure}[t]
		\centering
		\begin{tikzpicture}[node distance=1cm,scale=1.5]
			\input{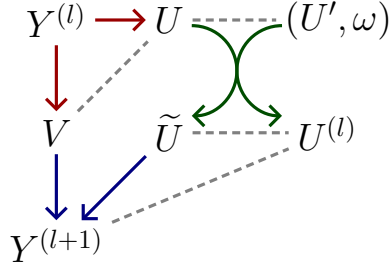}
		\end{tikzpicture}
		\caption{\small\textbf{EFR-SD via Factorization.} Arrows show functional relationships. Red arrows represent splitting $Y$ into $U$ and $V$ (the extraction step); 
			green arrows, exchanging information between $U$ and $U'$ to obtain $\wt{U}$ and $U^{(l)}$ (the filtration step); 
			blue arrows, generating $Y^{(l+1)}$ from $\wt{U}$ by blending with the auxiliary information $V$ (the reconstitution step). 
			Dashed gray lines indicate pairwise independence.} \label{F:independence}
	\end{figure}

	Each iteration $l = 1,\ldots,L$ of EFR-SD via factorization involves the following three steps, depicted in Figure \ref{F:independence}.
    
	\begin{enumerate}
		\item \textbf{Extraction:} We extract a statistic $U$ with information relevant to $\Hg$ from $Y^{(l)}$. We apply a {\em bijective factorization} function $\psi\left(Y^{(l)}\right)=(V,U)$ (in arbitrary probability spaces). $V$ can be interpreted as ancillary information that will be preserved for future SD iterations.
		\item \textbf{Filtration:} We add noise to split $U$ into new random variables $(\wt{U},U^{(l)})$ with an eye toward making $\wt{U}$ as close to $U$ as possible, while moving information relevant for inference into $U^{(l)}$. 
		More precisely, we start by sampling $(U',\omega)$ independent (of all other variables). 
		$U'$ will typically -- but need not -- have the same distribution as $U$, while $\omega$ is some random variable providing auxiliary randomization. 
		Then, we apply a deterministic {\em selection-filtration} function (depending on $\omega$), mapping the random variables $(U,U')$ to $(\wt{U},U^{(l)})$. 
		We elaborate on valid filters in Section \ref{sec:filtering}. 
		\item \textbf{Reconstitution:} Given $\psi$, we simply set $Y^{(l+1)}:= \psi^{-1}(V,\wt{U})$.
	\end{enumerate}
	
	\begin{theorem}
		\label{T:independentEFR}
		The EFR-SD via factorization procedure described above satisfies the conditions for generating a stable distillation if (1) the two components $\psi(Y)=(V,U)$ are independent, (2) $U$ and $\wt{U}$ have the same distribution, and (3) the two components $\wt{U}$ and $U^{(l)}$ are independent.
	\end{theorem}
	
	\noindent We complete the procedure by accumulating every $U^{(l)}$ into $U^\star = \left(U^{(l)}\right)_{l=1}^L$. Figure \ref{F:distdiagram} in Appendix \ref{sec:efrsd_diagram} illustrates how this accumulation into $U^\star$ and multiple EFR rounds fit together. As shown in Figure \ref{F:distdiagram}, after iterating EFR-SD $L$ times, we can also obtain a final $Y^{(L+1)}$ that is independent of the collection $\left(U^{(l)}\right)_{l=1}^{L}$. The by-product $Y^{(L+1)}$ may be used for further inference, though we will not specifically make use of it in this paper. 	
    
    Some knowledge of the null distribution of each $U^{(l)}$ is required in order to use it for inference. For the remainder of the paper, we will choose EFR-SD procedures that make each $U^{(l)}$ a vector of independent uniform random variables under $\Hg$. To help us do this, we will define a $\psi$ that makes each $U$ a vector of independent uniform random variables when $\Hg$ holds and randomly sample a vector $U'$ at the start of each filtration step with the same distribution. As we will show in the next section, in the context of the OLS model, this construction will allow us to interpret each entry of $U^\star$ as an independent p-value associated with a given predictor.
	
	\subsection{Simple OLS Example} \label{sec:ols_example}
	
	Here we will introduce a simple example of EFR-SD via factorization, where we sequentially estimate each $\beta_j$ in the OLS model \eqref{model}. 
	In Appendix \ref{sec:dimensionality}, we consider a more general case where our $p$ predictors are grouped into $L < p$ ``layers,'' with each layer of predictors corresponding to one iteration of SD. 
	However, throughout this section, we will assume that we have $L=p$ ``atomic'' layers: that each iteration of SD will involve only one predictor $X_{\cdot j}$. Without loss of generality, for ease of notation, in this section we assume that the columns of $X$ are permuted so that $X_{\cdot j}$ is considered in the $j^{th}$ SD iteration. 
	
	The classic $F$-statistic for testing $\Hg : \beta_j = 0$ for a single $\beta_j$ is 
	\begin{equation}
		\label{classic_t}
		T^2_j = \frac{W^2_j}{(\omega - W_j^2) / (n-q-1)}
	\end{equation}
	
	\noindent where $W_j = Y^\top \wt{X}_j$,  and $\wt{X}_j = P^\bot_A X_{\cdot j}  / \left|\left|P^\bot_A X_{\cdot j}\right|\right|_2$,
	and $\omega = \left|\left|P^\bot_{A} Y\right|\right|_2^2$. 
	This statistic pivots out the unknown parameters $\alpha$ and $\sigma$. 
	Under $\Hg : \beta_j = 0$, $T_j^2$ has an $F$-distribution with 1 and $n-q-1$ degrees of freedom, so mapping $T^2_j$ through the corresponding complementary CDF yields a p-value $U_j$. The key to applying SD in this context is to notice that this extraction of $U_j$ from $Y^{(j)}$ may be inverted while maintaining stability of distribution and the required independence assumptions. 
	We can write the extraction of the statistic $U_j$ from $Y^{(j)}$ as a component of an bijective factorization $\psi_j$ of $Y^{(j)}$ into four independent pieces as follows. To avoid the technical issues raised by the possibility that $P_{A,X_{\cdot j}}^\bot Y = 0$, we exclude the subspace of $\R^n$ spanned by the columns of $(A,X_{\cdot j})$ from our proposed map $\psi_j$, a set of measure zero. Specifically, let $\mathfrak{N} = \left\{ x \in \R^n: P_{A,X_{\cdot j}}^\bot x \neq 0 \right\}$ and  $\mathfrak{S} = \left\{ x \in \mathfrak{N}: \|x\|=1\right\}$.
	
	\begin{prop}
		\label{prop_1_var}
		The map $\psi_j: \mathfrak{N} \to  \R^q  \times (0,1] \times \{-1,0,+1\} \times \R_+\times \mathfrak{S}$ given by \\
		$\psi_j: Y \mapsto \left(P_A Y, \ U_j, \ \sgn\left(T_j\right), \ \omega , \ P^\bot_{A,X_{\cdot j}} Y / \left|\left|P^\bot_{A,X_{\cdot j}} Y\right|\right|_2 \right)$ is invertible. If \  $Y \sim N(A\alpha,\sigma^2 I_n)$ for some $\alpha \in \mathbb{R}^q$ and $\sigma^2 \in \mathbb{R}_+$, then the five components of $\psi_j(Y)$ are mutually independent.
	\end{prop} 
	
	\noindent By Theorem \ref{T:independentEFR}, the family of maps $\psi_j$ may be used to generate a SD process. Recall that while EFR-SD via factorization requires the existence of such an explicit $\psi_j$, SD in general does not. Since we expect each p-value $U_j$ obtained in this section to be $\textrm{Unif}(0,1)$ under $\Hg$, we sample a new $U' \sim \textrm{Unif}(0,1)$ at each iteration in accordance with the filtration step from Section \ref{sec:independentEFR}.
	
	Now we are ready to introduce an EFR-SD procedure for the OLS model that uses a trivial filter where the extracted $U_j$ is always stored in $U^{(j)}$ and noise is injected into $Y$ at every step; we call this the {\em na\"ive SD procedure}. 
	We start with the extraction step, calculating $\psi_j(Y^{(j)})$. 
	Then, in the filtration step, we simply set $U^{(j)} = U_j$ and $\wt{U} =U' \sim \mathrm{Unif}(0,1)$. Finally, in the reconstitution step, we invert $\psi_j$ to obtain $Y^{(j+1)} = \psi_j^{-1}\left(P_AY, \ \wt{U}, \ \sgn\left(T_j\right), \ \omega, \ P^\bot_{A,X_{\cdot j}} Y / \left|\left|P^\bot_{A,X_{\cdot j}} Y\right|\right|_2 \right)$.
	
	This procedure can be implemented algorithmically in a somewhat more simplified form as follows. In the extraction step, we calculate $U_j$ via $T_j^2$. In the reconstitution step, we map $\wt{U}$ through the inverse complementary CDF of the appropriate $F$-distribution to otain $\wt{T}_j^2$. We then obtain $\wt{W}_j = \sgn\left(T_j\right) \sqrt{\frac{\omega \wt{T}^2_j}{(n-q-1) + \wt{T}^2_j}}$. Finally we set
	\begin{equation}
		\label{eq:simple_inversion}
		Y^{(j+1)} = P_A Y^{(j)} +  \gamma \left(P^\bot_A Y^{(j)} - W_j \wt{X}_j \right) + \wt{W}_j \wt{X}_j \quad \mathrm{where} \quad \gamma = \sqrt{\frac{\omega - \wt{W}_j^2}{\omega - W_j^2}}.
	\end{equation}
	Under $\Hg$, Proposition \ref{prop_1_var} guarantees stability --- $Y^{(j+1)} \sim N(A\alpha,\sigma I_n)$ --- and that $Y^{(j+1)}$ is independent of $U^{(j)}$. 
	From \eqref{eq:simple_inversion}, the change in $Y$ induced by this procedure can be written as
	\begin{equation}
		\left|\left|Y^{(j+1)} - Y^{(j)} \right|\right|^2_2 = \underbrace{\left(\wt{W}_j - \gamma W_j\right)^2}_{\textrm{shifting}} + \underbrace{\omega (\gamma-1)^2}_{\textrm{rescaling}},
	\end{equation}
	which neatly shows the noise coming from shifting $Y^{(j)}$ along $\wt{X}_j$ and the noise coming from rescaling $Y^{(j)}$. 
	While this yields a valid SD procedure that can be iterated, it is not powerful because of the relatively large amount of noise injected into $Y^{(j)}$ at each iteration. 
	This motivates the use of a non-trivial filter, as explored in the next section.

	\section{Filtering}\label{sec:filtering}

    We open this section by showing how a simple filter can rescue the na\"ive SD procedure for the OLS model \eqref{model} introduced in Section \ref{sec:ols_example}. After investigating the power provided by this simple filter, we introduce our more general filtering framework.
    
	\subsection{The Simple Quantile Filter} \label{sec:quantile}
	
	Here revisit our na\"ive SD procedure for the OLS model \eqref{model}, adopting the notation introduced in Section \ref{sec:ols_example}. Rather than applying the trivial filter used in Section \ref{sec:ols_example}, consider equipping the same EFR-SD procedure with the Simple Quantile Filter, for some fixed quantile $t \in (0,1)$, shown in Algorithm \ref{alg:simple_quantile_filter}. We will refer to this as the \emph{filtered SD procedure}. 
	\spacingset{1}
	\begin{algorithm}[!ht] 
		\caption{Simple Quantile Filter}\label{alg:simple_quantile_filter}
		\textbf{Input:}  $t,\ U_j, \ U' \quad \quad \quad \quad \quad \quad \quad \quad \quad \quad \quad \quad \quad \quad \quad \quad \quad \quad \quad \quad \quad \quad \quad \quad \quad \quad \quad \quad \quad \quad \quad \quad \quad \quad   \quad \quad \quad$\\ 
		\textbf{Output:} $U^{(j)},\ \wt{U} \ \ \quad \quad \quad \quad \quad \quad \quad \quad \quad \quad \quad \quad \quad \quad \quad \quad \quad \quad \quad \quad \quad \quad \quad \quad \quad \quad \quad \quad \quad \quad \quad \quad \quad \quad \quad  \quad $ 
		\begin{algorithmic}
			\Require $t,\ U_j, \ U' \in (0,1)$
			\If{$U_j < t$}
			\State $\wt{U} \gets U'$ 
			\State $U^{(j)} \gets U_j$
			\Else
			\If{$U' > t$}
			\State $\wt{U} \gets U_j$
			\State $U^{(j)} \gets U'$
			\Else  
			\State $\wt{U} \gets t \cdot g_t\left(U_j\right)$ \Comment{where $g_t(x) = (x-t) / (1-t)$}
			\State $U^{(j)} \gets g^{-1}_t \left(U' / t \right)$ 
			\EndIf 
			\EndIf \\
			
			\Return $U^{(j)},\ \wt{U}$
		\end{algorithmic}
	\end{algorithm}
	\spacingset{1.9}
	
    \noindent In Algorithm \ref{alg:simple_quantile_filter}, if $U_j > t$ and $U' > t$, which is highly likely under $\Hg$ for small $t$, then $U_j$ is unchanged and the filter simply returns $\wt{U} = U_j$, guaranteeing that $\wt{W}_{j} = W_{j}$ and $Y^{(j+1)} = Y^{(j)}$. 
	More explicitly, under $\Hg$, this filter guarantees that  $\p{Y^{(j+1)} = Y^{(j)}} \ge (1-t)^2$ and   $\mathbb{E}\left[\left|\left|Y^{(j+1)} - Y^{(j)} \right|\right|^2_2\right] \leq C \cdot t$ for some $C > 0$. 
    
    Setting $t$ lower not only reduces the expected noise introduced in each iteration, but also reduces the risk of proxy predictors (predictors correlated with the active predictors) truncating the signal driven by the active predictors. On the other hand, if the filtering threshold $t$ is set too low, small outlying $U_j$ corresponding to important active predictors may simply be missed (omitted from $U^\star$). We revisit the choice of $t$ in our simulation studies presented in Section \ref{sec:power_sims}. 
    
    \subsection{Power in the case of uniformly scattered null predictors} \label{sec:power2}
	\newcommand{\hmt}{\hat\mu^\top}

    Here analyze the power of our filtered SD procedure for the OLS model under the alternative hypothesis where $Y \sim N\left(\mu, \sigma I\right)$ for some arbitrary $\mu \in \mathbb{R}^n \setminus 0$. While it is difficult to assess the power of filtered SD for any fixed design matrix $X$, here we show that the procedure is consistent and provide a non-asymptotic bound on the power it achieves in a special case where the ``null predictors'' are assumed to represent random directions. 

    More explicitly, say we have $K$ ``true predictors'' and $p-K$ ``null predictors''.
	Assume that the null predictors $X_{\cdot j}$ (for $j\in\{1,\dots,p\}\setminus \{j_1,\dots,j_K\}$) are i.i.d.\ vectors uniformly distributed on $S^{n-1}$, the unit sphere in $\R^n$.
	Assume each true predictor $X_{\cdot j}$ has projection $b$ onto $\mu$ for a fixed $b\in (0,1)$; that is, $\mu^\top X_{\cdot j} =b\|\mu\|$.
	The projections of each true predictor orthogonal to $\mu$ are also uniform on the sphere, now with radius $\sqrt{1-b^2}$ in the $(n-2)$-dimensional space orthogonal to $\mu$, and also independent of each other and all other predictors.

    While we do not expect these assumptions to be literally true in practice, this set-up is instructive because it captures two essential features of the sparse signal setting where existing methods struggle. First, there are a very large number of almost-null predictors, in the sense that their alignment with the true mean $\mu$ is purely coincidental --- represented here by the uniform direction --- but not zero. Second, the signal is divided up among multiple true predictors.

    Let $k$ be the anticipated number of true predictors, inspiring a test at level $\alpha$ with a simple combination statistic that is $\xi=\max\sum_{l=1}^k -\log U_{j_l}$,
	where the maximum is over all choices of $k$ predictors.
	Let $\xi_\alpha$ be the critical value for the test, which is approximately $p\log\frac{p}{k} -\log\alpha$.
    In Appendix \ref{sec:power_appendix}, we derive a finite-sample lower bound on the power that goes to 1 exponentially in $\sqrt{n}$, with a rate that depends essentially on the ratio $\mu_1^2/\xi_\alpha$, where $\mu_1^2$ is the proportion of $\|Y\|^2$ contributed by $\mu$ --- essentially $\|\mu\|^2/n$. Our bound implies the following asymptotic result.

\begin{corollary} \label{C:power}
Suppose $Y = \mu_n + \sigma\epsilon$, where $\epsilon\sim N(0,I_n)$. 
For each $n$ we perform the SD procedure on $p$ predictors that satisfy the above conditions (in particular, that there are $K$ active predictors, with correlation $b$ to $\mu$) at a fixed level $\alpha$.
There are constants $C>0$ and $M$, depending on $b$, $t$, and $\alpha$ (but not $\mu$, $p$, or $n$) such that, for $\|\mu\|>M k \log p$,
\begin{equation*}
 \log(1-\operatorname{Power} ) \le
   -C\min(\sqrt{n}+tp, \|\mu\|^2).
\end{equation*}
\end{corollary}

	\subsection{Power Simulations} \label{sec:power_sims}
	Here we apply SD to data simulated under the OLS model \eqref{model} where $p=10,000$. First we consider the $n > p$ case, setting $n=100,000$. We simulate $X$ so that the columns follow a block covariance structure based on clusters of size 10. We consider three levels of dependence within the blocks of predictors in terms of the proportion of variance shared, $r^2 \in \left\{0.2, 0.5, 0.8\right\}$. For each simulation, we select a desired number of active predictors $a \in \left\{4,16\right\}$ uniformly at random from $[p]$. Given this active set $\mathcal{A}$ we select $\beta$, while aiming to distribute the observed effects across active predictors as evenly as possible for a fixed underlying ``signal strength'' $s$. This $s$ is constructed to be approximately the $-\log_{10}$ p-value that an oracle ANOVA model, which ``knows'' the active predictors, would return in a given simulation. For each combination of $a \in \{4,16\}$, $r^2 \in \left\{0.2, 0.5, 0.8\right\}$, and $s \in \{0,2,4,6,\ldots,200\}$, we simulate 2,000 independent pairs $\left(Y,X\right)$, each with a different $\mathcal{A}$. Here we run our filtered SD procedure separately across a range of filter thresholds $t$, taking the Bonferroni corrected minimum among the resulting p-values as our final p-value. These thresholds are chosen to maximize power given the required size of the test $\alpha$ and account for the unknown number of active predictors $a$. See Section \ref{sec:power_sim_details} for further simulation details. 
    
\begin{figure}[!ht]
		\centering
		\includegraphics[width=0.95\linewidth]{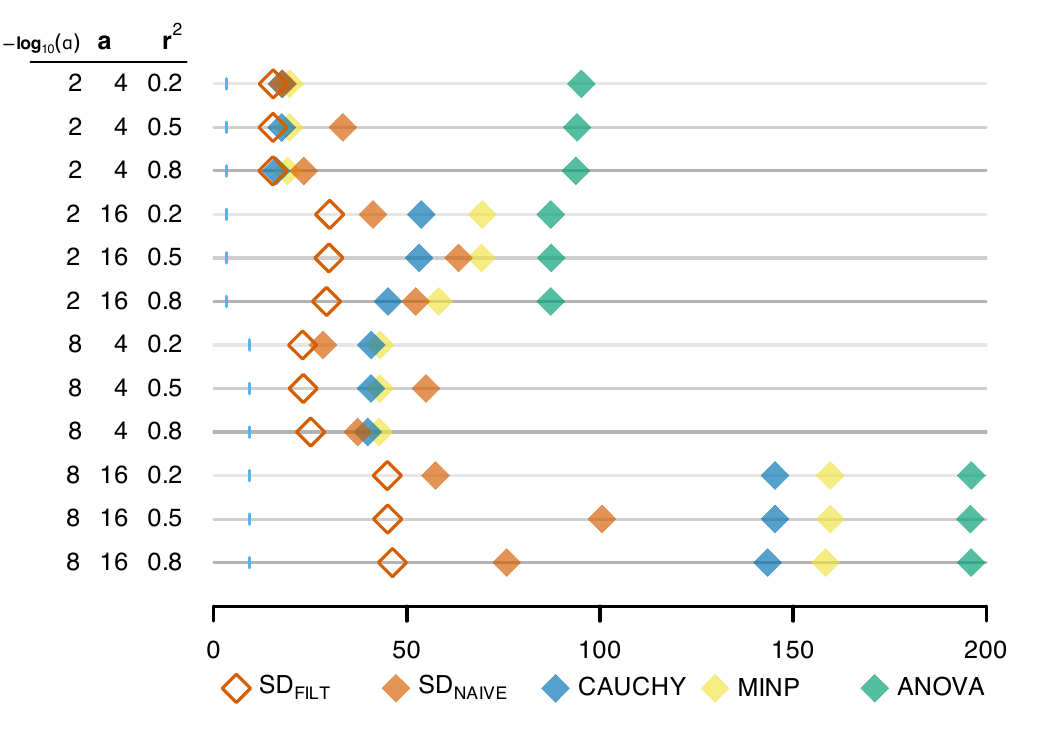}
		\caption{\small \textbf{Power in the $n>p$ case.} Dotplot of signal strength ($s \ \approx$\ oracle $-\log_{10}$ p-value) required to achieve 80\% power under various conditions. Here, $p = 10,000$ and $n=100,000$. The columns of the key on the left list the size of the test ($10^{-2}$ or $10^{-8}$), the number of active predictors ($a$), and the squared correlation between predictors within each block/cluster $r^2$ respectively. The light blue tick marks on each line indicate the effect size at which the oracle model achieves 80\% power. The absence of a diamond on any line indicates that it is beyond the range of the plot (signal strength $>$ 200).}
		\label{dense_large_n_sims}
	\end{figure}

     We compare our filtered SD  and na\"ive SD approaches with the Cauchy Combination Test \cite{liu2020cauchy} (CAUCHY), the Bonferroni-adjusted minimum marginal p-value (MINP), and ANOVA. This yields the power curves shown in Section \ref{sec:power_curves}, where each plotted point is based on 2,000 replicates. We summarize the resulting power curves in Figure \ref{dense_large_n_sims}, where we plot the estimated signal strength $s$ required to achieve 80\% power. 
	
	Figure \ref{dense_large_n_sims} shows that filtered SD dominates standard methods in all settings and significantly improves on the non-filtered, na\"ive SD method. When there are 4 active predictors and the size of the test is $10^{-2}$ the power gain is minimal; when there are 16 active predictors, considerable. When the size of the test is $10^{-8}$, the power gain available through filtered SD is particularly notable, requiring less than half the signal of the Cauchy method in the case of 16 active predictors. We also observe that the power gain available through filtered SD is typically higher when $r^2$ is lower. We replicated the experiments displayed in Figure \ref{dense_large_n_sims} while setting $n = 1,000$. The results in this $n < p$ case are shown in Figure \ref{fig:dense_small_n_sims} of Appendix \ref{sec:dense_small_n_sims}, where we see that filtered SD continues to dominate standard methods. 
    
    The Cauchy method (using $p$ BLAS-parallelized projections) and filtered SD scale as $\mathcal{O}(np)$. We not only observed similar scaling for the two algorithms in $n$, but also that the actual run time of SD was very close to that of the Cauchy method for large $n$. See Appendix \ref{sec:scalability} for details.

    \subsection{General Filtering} \label{sec:generalfiltering}
    Here we provide a general framework for constructing valid filters, returning to the general notation for SD processes used in Section \ref{sec:sd_def}. At a high level, filtration exchanges information between some $U$ and some simulated $U'$. The choice of which information to exchange can be made to maximize the power while minimizing the distortion of $Y$. Here we assume that we have selected an SD procedure such that, during SD iteration $l$, both $U$ and $U'$ are vectors of independent uniform random variables with length $p_l$. Let $r : \left\{1,\ldots,p_l\right\} \rightarrow \left\{1,\ldots,p_l\right\}$ be the ordering permutation of $U$, so that $U_{r(1)} \leq U_{r(2)} \leq \ldots \leq U_{r(p_l)}$.
	Let $T = \left(U_{r(1)}, U_{r(2)}, \ldots, U_{r(p_l)} \right)$ be the order statistic of $U$, and let $T'$ be the order statistic of $p_l$ simulated independent standard uniform random variables, $U'$.
	
	\newcommand{\pstar}{}
	Let $\varphi : [0,1]^{p_l} \rightarrow \mathcal{P} \times \mathcal{Q} \times \prod\limits_{i=1}^m \mathcal{R}_i$ be an invertible \emph{filtration function} where $\mathcal{P}$, $\mathcal{Q}$, and $\mathcal{R}_1,\ldots,\mathcal{R}_m$ are arbitrary state spaces.
	We require that $\varphi$ be chosen such that $\left(P^{\pstar},Q^{\pstar},R^{\pstar}_1,\ldots,R^{\pstar}_m\right) = \varphi\left(Z\right)$ are mutually independent when $Z$ is the order statistic of $p_l$ independent standard uniform random variables.
	Let $\left(P,Q,R_1,\ldots,R_m\right) = \varphi\left(T\right)$ and $\left(P',Q',R'_1,\ldots,R'_m\right) = \varphi\left(T'\right)$.
	Filtration proceeds according to these simple rules: $P$ and $P'$ are never exchanged; $Q$ and $Q'$ are always exchanged; whether components $R_i$ and $R'_i$ are exchanged is determined by $Q$.
	That is, we create two new versions of the data $\wt{T}$ and $T$, mixing $\varphi(T)$ and $\varphi(T')$:
	\begin{align}
		\wt{T} &= \varphi^{-1}\left(P,Q',\wt{R}_1,\dots,\wt{R}_m \right), \\
		T &= \varphi^{-1}\left(P',Q,R^\star_1,\ldots,R^\star_m \right),
	\end{align}
	where
	\begin{equation}
		(R^\star_i , \wt{R}_i) := \begin{cases}
			(R_i, R'_i) & \text{if } \S\left(Q\right)_i =1\\
			(R'_i, R_i) & \text{if } \S\left(Q\right)_i =0\\
		\end{cases}
	\end{equation}
	for an arbitrary \emph{selection function} $\S: \mathcal{Q} \rightarrow \left\{0,1\right\}^m$.
	
	We set $\wt{U} = \left( \wt{T}_{r^{-1}(1)},  \wt{T}_{r^{-1}(2)}, \ldots , \wt{T}_{r^{-1}(p_l)}\right)$ and $U^{(\ell)} = \left( T_{r^{-1}(1)},  T_{r^{-1}(2)}, \ldots , T_{r^{-1}(p_l)}\right)$. Given the mutual independence among the components $\left(P^{\pstar},Q^{\pstar},R^{\pstar}_1,\ldots,R^{\pstar}_m\right)$ guaranteed by $\varphi$, this swapping procedure ensures that $\wt{U} \indep U^{(l)}$. From here $U^{(l)}$ is stored and $\wt{U}$ is used to generate $Y^{(l + 1)}$ in the reconstitution step. Note that $\varphi$ and $\S$ can vary between SD iterations.
	
	A simple example of $\varphi$ involves exchanging all components of $U$ below a specified quantile cutoff $t$ -- a generalization of the simple quantile filter in Algorithm \ref{alg:simple_quantile_filter}. This type of filter is useful in cases where we have multiple predictors in each distillation iteration (layer), as further described in Appendix \ref{sec:dimensionality}. 
    Let $g_t(x) = (x-t) / (1-t)$ and let $\rho$ be the \textit{R\'enyi transformation}, described in Appendix \ref{sec:renyi}, that transforms ordered independent $\textrm{Unif}(0,1)$ random variables into independent unit exponential random variables.
	
    \begin{definition} Quantile Filter \\
		Given a quantile cutoff $t \in (0,1)$, a quantile filter uses filtration function $\varphi_t(T) = (Q,R)$ where
		\begin{align}\label{E:quantile}
			Q &= \#\{j\in [p_l]\, : \, T_j \le t\} \\
			R &= \left(\rho\left(T_1/t,\ldots,T_Q/t \right),\rho\left(g_t\left(T_{Q+1}\right),\ldots,g_t\left(T_{p_l}\right) \right) \right)
		\end{align}
		and selection function $\S\left(Q\right)_i =\mathbf{1} \left\{ i \leq Q \right\}$.
	\end{definition}
	\noindent This quantile filter is a generalization of Algorithm \ref{alg:simple_quantile_filter} and allows us to apply quantile filtering to distillation procedures where there are multiple predictors included in each iteration. Using this more general filtering framework, we introduce another class of filters based on exchanging the smallest $k$ order statistics in Appendix \ref{sec:smallest_k_filtering}.

\section{EFR-SD via Coupling} \label{sec:generalEFR} 

The EFR via factorization framework introduced in Section \ref{sec:independentEFR} proposes decomposing the data into two independent (under the null hypothesis) pieces $(U,V)$ via a bijective factorization function $\psi$. While such a decomposition is always available in principle, see Appendix \ref{sec:general_decomposition}, finding an explicit $\psi$ for any given model is challenging: we only easily obtained it in the case of the OLS model based on the independence of orthogonal Gaussian projections. The essential role of $\psi$ is in the reconstitution step where it allows us to construct a new $\wt{Y}$ given $\wt{U}$. Building on this observation, here we propose a more general EFR-SD framework where we sample a new $\wt{Y}$ given $\wt{U}$ while coupling $\wt{Y}$ to the initial data $Y^{(l)}$. The tighter the coupling, the less noise injected into $\wt{Y}$. Below we show that this approach yields valid SD procedures. We provide an explicit example for the case of logistic regression in Section \ref{sec:LR_example}.

\subsection{The Coupling Framework}
	Let $\pi_u(\diff y)$ be a regular conditional distribution for $Y$ conditioned on $U$,
	and define a coupling of $\pi_u(\diff y)$ and $\pi_{\tilde u}(\diff \tilde y)$.
	The coupling may be represented as a joint distribution $\pi_{u,\tilde{u}}(\diff y, \diff \tilde y)$, where the marginal distributions are  $\pi_u(\diff y)$ and $\pi_{\tilde u}(\diff \tilde y)$,
	or as a regular conditional distribution, which is a kernel $\pi_{\tilde{u},y}( \diff \tilde{y})$.
	Similarly, we may think of the filtration step as being determined by a joint distribution $\f(\diff u, \diff \tilde u, \diff u^{(l)})$ that couples $U=U(Y^{(l)})$ to the new independent pair $\wt{U},U^{(l)}$. Here we use $q(\diff u)$ to denote the marginal distribution of $U$.
	
    Each iteration of EFR-SD via coupling is carried out as follows:
	\begin{enumerate}
		\item \textbf{Extraction:} Starting from $Y^{(l)}$, we calculate $U$. 
        \item \textbf{Filtration:} Given $U$, we sample the independent pair $(\wt{U},U^{(l)})$ using $\f$.
		\item \textbf{Reconstitution}: From the kernel $\pi_{\wt{U},Y^{(l)}}$ we generate a new $Y^{(l+1)}$.
	\end{enumerate}

	\begin{figure}[t]
		\centering
		\begin{tikzpicture}[node distance=1cm,scale=1.5]
			\input{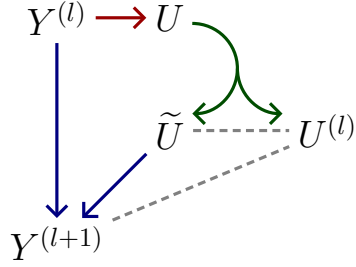}
		\end{tikzpicture}
		\caption{\small\textbf{General EFR-SD.} Arrows show functional relationships. The red arrow represents the single mapping of $Y$ to $U$ (the extraction step); 
			green arrows, the splitting of $U$ into $\wt{U}$ and $U^{(\ell)}$ (the filtration step); 
			blue arrows, generating $Y^{(l+1)}$ from $\wt{U}$ and the residual information in $Y^{(l)}$ (the reconstitution step) through the coupling $\pi$. 
			Dashed gray lines indicate the independence of variables.} \label{F:generalEFR}
	\end{figure}
	
	\begin{theorem}
		\label{T:generalEFR}
		The general EFR-SD procedure described above satisfies the conditions for generating a stable distillation if the following hold:
		\begin{itemize}
			\item  The joint distribution $\pi_{u,\tilde{u}}(\diff y, \diff \tilde y)$ has marginal distributions $\pi_u(\diff y)$ and $\pi_{\tilde u}(\diff \tilde y)$ for every choice of $u$ and $\tilde u$;
			\item The joint distribution $\f(\diff u, \diff \tilde u, \diff u^{(l)})$ has marginal distribution $q$ in each component, and makes $\wt{U}$ and $U^{(l)}$ independent.
		\end{itemize}
	\end{theorem}
	
	Note that if we have an invertible splitting $\psi(Y)=(U,V)$, and a filtering function $\phi(u,u',\omega)$, we may define kernels
	\begin{align*}
		\pi_{u,\tilde{u}} f(y) g(\tilde y)& := \E[ f(\psi^{-1}(u,V)) g(\psi^{-1}(\tilde u,V)) ] ,\\
		\f_u f(\tilde u, u^{(l)}) &:= \E[ f(\phi(u,U', \omega))]
	\end{align*}
	for bounded measurable test functions $f: \Y\times \Y\to \R$ and $g: \mathcal{U}\times \mathcal{U} \to \R$.
	If the functions satisfy the conditions of Theorem \ref{T:independentEFR} then the kernels satisfy the conditions of Theorem \ref{T:generalEFR}; 
	hence the EFR-SD via factorization procedure is a special case of the coupling version.
	Theorem \ref{T:independentEFR} could be a corollary of Theorem \ref{T:generalEFR}, but we present a separate proof for expository reasons. While this coupling approach may appear unwieldy, it can produce relatively simple procedures: the SD procedure we provide for logistic regression in Section \ref{sec:LR_example}.

	\subsection{Distilling the logistic regression model} \label{sec:LR_example}
	While many inference problems can be coaxed into the form of testing a Gaussian model, this is often not possible for categorical outcomes.
	To illustrate the flexibility of the EFR-SD via coupling framework, here we present an SD procedure for logistic regression. For some binary data $Y \in \{0,1\}^n$, consider testing $\Hg : \beta \equiv 0$ under the logistic regression model $Y_i \sim \mathrm{Bern}\left( \operatorname{logistic}(A_{i \cdot}\alpha + X_{i \cdot}\beta)\right)$.
	We begin by fitting the background model $Y_i \sim \mathrm{Bern}\left( \operatorname{logistic}(A_{i \cdot}\alpha)\right)$, to obtain background estimates $\hat p_i$ for $\P(Y_i=1)$. As we did for the OLS model, for ease of notation, we assume that the columns of $X$ are permuted so that $X_{\cdot j}$ is considered in the $j^{th}$ SD iteration.
	During each extraction step, we use the score test statistic 
	\begin{equation}
		\frac{ ((Y - \hat p)^\top X_{\cdot j})^2}{\sum_{i=1}^n \hat p_i (1-\hat p_i)X_{ij}^2} .
	\end{equation}
	After shifting and scaling, this score statistic is equivalent to $S_n := \sum_{i=1}^n Y_i X_{ij}$,  which has a Poisson binomial distribution with parameters $\hat{p} := (\hat p_1,\dots,\hat p_n)$. 
	This is a discrete distribution, with corresponding cdf $G_{\hat{p}}$, so we cannot directly apply it to $S_n$ to obtain a uniformly distributed p-value.
	We can add auxiliary randomization to fill in the gaps. 
	Define $\omega$ to be uniformly distributed on $[0,1]$ and independent of all other random variables.
	If the possible values of $S_n$ are $s_1<s_2< \cdots < s_m$, and setting $s_0:=-\infty$, we define $W := \omega G_{\hat{p}}(s_i) + (1-\omega) G_{\hat{p}}(s_{i-1}) \text{ when } S_n = s_i$.
	This is analogous to the use of randomized residuals \cite{RQR}. This $W$ has uniform distribution on $[0,1]$ when $Y$ is a sample from the Poisson binomial distribution with parameters $\hat{p}$. Decomposing $W$ to obtain a ``two-sided p-value,'' as we did for the OLS model, we have $\mathbf{1}\left\{W\leq 0.5\right\}$ and $U:= 2 \min(W,1-W)$. After filtering we have two random variables $\wt{U}$ and $U^{(j)}$ that have the same uniform distribution. We set $\wt{W} = \left(\wt{U}/2\right)^{\mathbf{1}\left\{W\leq 0.5\right\}}\left(1-\wt{U}/2\right)^{1-\mathbf{1}\left\{W\leq 0.5\right\}} $.
	
	To complete the EFR procedure we need a coupling between the distribution $\pi_w$ of $(Y_1,\dots,Y_n)$ conditioned on $Y^\top X_{\cdot j}= r := G_{\hat{p}}^{-1}(W)$, and the distribution $\pi_{\tilde w}$ of $(Y_1,\dots,Y_n)$ conditioned on $\wt{Y}^\top X_{\cdot j}=\tilde r :=G_{\hat{p}}^{-1}(\wt{W})$.
	One way to do this is to view the partial sums $S_k := \sum_{i=1}^k Y_i X_{ij}$ as a random walk. See Appendix \ref{sec:logistic_regression_coupling_equations} for details. This approach automatically returns $Y^{(j+1)}=Y^{(j)}$ whenever $r = \tilde{r}$, introducing no additional noise into the data. This happens whenever $\wt{U}=U$, which should be a frequent occurrence since we expect the large majority of predictors to fail the filter. 
    Another major advantage of this approach is its computational efficiency.
    While the probabilities required at each step may seem to require sums over exponentially many (in $n$) terms, they may be computed rapidly via the Fast Fourier Transform, as explained in \cite{BZB18}. 
	
    A weakness of this approach is that increased diversity among the entries of each $X_{\cdot j}$ can reduce statistical efficiency. While our coupling procedure is valid for any $X_{\cdot j}$, the more various the entries of $X_{\cdot j}$ become, the more $Y^\top X_{\cdot j}$ will constrain the entire sequence $Y$, and so, the more noise the procedure will inject into $Y^{(j+1)}$ when $\wt{U}\neq U$.
	In the extreme case, where $X_{\cdot j}$ contains distinct real entries (at arbitrary precision), each possible sum will be obtainable in exactly one way, and the ``coupling'' will be singular.
	In other words, the ``outcome data'' $Y^{(j)}$ will rapidly degenerate to mere noise, and power will be essentially zero. 
	This does not reflect a flaw in the coupling approach, but rather a requirement born out of the discreteness of $Y$ and the independence properties required to produce a valid SD process. 
	
	We can avoid degeneracy and maintain statistical power in practice if most of the possible values of $Y^\top X_{\cdot j}$ are achievable via many random walk paths. We will have this redundancy among the paths when $X$ consists of small integers. A common example includes applications where $X \in \{0,1\}^{n\times p}$; each predictor may correspond to a different (potentially overlapping) treatment condition. Another example includes genomics applications where predictors correspond to variants, with $X_{ij}$ representing the genotype --- 0, 1, or 2 copies --- for individual $i$ at variant $j$. These examples suggest a more general, approximate approach: scaling and shifting each $X_{\cdot j}$ so that it may be approximated with a new predictor that takes values among small integers. This may achieved computationally by casting each $X_{\cdot j}$ to a low-precision (eg: single-byte) representation. An alternative approach is to treat the entries $X_{ij}$ as being themselves subject to random noise, an idea that we briefly outline in Appendix \ref{sec:avoiding_degeneracy_for_logistic_regression}.

	\section{Connections to the Multiple Testing Literature} \label{sec:literature}
	Here we situate SD within the post-selection inference literature. Many of the approaches in the multiple testing literature we discuss below target the much more challenging task of variable selection, with FDR or FWER control. We do not take on variable selection in this paper -- the primary aim in post-selection inference -- so it would be unfair for us to draw any direct comparisons to our present work. We provide the following high-level discussion to give some intuition for the strengths and weaknesses of SD as a foundation for inference within and beyond the context of testing $\Hg$. 
	
	The idea of injecting noise into $Y$ to break the dependence between predictors may seem unorthodox, or even ill-advised, but is well established.
	Inferences from Monte Carlo methods, and particularly Bayesian inferences based on MCMC, are obviously stochastic, as is cross-validation.
	The same applies to knockoff procedures, such as the fixed-X knockoff filter \cite{barber2015}, which have become popular tools in variable selection, particularly in genomics.
	\cite{whiteout} described how the fixed-X knockoff filter can be rewritten as a whitening procedure that adds noise to effect size estimators $\hat\beta_j$ to make them independent.  The greater the dependence between predictors, the more noise is required.
	This whitening representation makes explicit the fact that, with fixed-X knockoffs, noise must be introduced to decouple the $\hat\beta_j$ {\emph{a priori}} -- without any information about $Y$. Although model-X knockoffs inject noise into the variable selection procedure by simulating knockoff predictors rather than knockoff outcomes \cite{candes2018}, the noise is still introduced {\emph{a priori}}. 
	
	In contrast, SD introduces noise \emph{a posteriori}, after observing relevant information from $Y$. Since we are assuming that the number of non-null hypotheses is small, during the filtration step we can substantially limit the information exchanged between $U$ and $U'$ to the most extreme observations.
	While this makes use of partial information from $Y$, to prioritize where to inject noise, the filter substantially reduces the amount of noise injected back into $Y$ at the end of each round of distillation via $\wt{U}$. As we saw in Section \ref{sec:ols_example}, sometimes an iteration of distillation can even return $Y$ completely unaltered. It is this \emph{a posteriori} approach that ultimately conserves power across iterations and the overall power of SD against sparse alternatives.
	
	While SD is also an interactive procedure, it is fundamentally different from recently-proposed iterative multiple-testing methods based on masking \cite{lei2018adapt, fithian2020, duan2020familywise}. These approaches iterate at the level of p-values rather than the data $Y$ itself. SD iteratively updates the data $Y$ itself, requiring a factorization $\psi$, or more generally, a coupling based on the assumed model $F_\theta$. As described above, the intended benefit of working ``closer to the data'' is more parsimonious noise injection. We do not compare to the very creative masking-and-martingale-based global null testing procedure developed by Duan et al. \cite{duan2020interactive} because that procedure requires p-values to be independent under the null hypothesis. However, p-values in $U^\star$ returned by SD could be used as input to that procedure.
	
	A key downside to SD's iterative approach is that it does not treat all hypotheses exchangeably: in regression models, predictors included in early layers will be tested against an outcome vector that is closer to the initial vector $Y$. However, the power consequences of this are mitigated by the filtration step. This sequential nature of SD has another key side-effect: when there is a strong effect harbored by a cluster of tightly correlated predictors, SD will ``extract'' the signal from whichever predictor is in the earliest layer as a representative and use it against $\Hg$. While this is an advantage from the perspective of global null testing, it makes FDR-style guarantees for variable selection based on $U^\star$ difficult. Any predictor $X_k$ that is strongly correlated with a true active predictor $X_j$ may act as a proxy if $X_k$ appears in an earlier layer than $X_j$. Due to their mandate to control FDR, methods with FDR guarantees tend to entirely ignore these clusters \cite{candes2018, wang2020simple}. The more relaxed FDR framework defended in \cite{barber2019} for settings like genomics, where a proxy predictor may be deemed acceptable in lieu of identifying the true active predictor, may provide a path towards SD-based variable selection procedures.
	
	The sequential nature of SD makes it similar in spirit to sequential ``data-carving'' methods for variable selection, built around the LASSO, which also tend to select one representative predictor from each cluster when active predictors are embedded in clusters of proxy predictors \cite{fithian2014optimal, lee2016exact}. However, the two-stage nature of these data-carving methods -- where part of the information in the data is used to select variables and the remaining information is used for inference -- has made it difficult for them to achieve sufficient power to be useful in practice \cite{tian2018selective, liu2018more}. Since, under this framework, the underlying distribution of $Y$ is ever changing as more variables are selected and those selection events are conditioned on, data-carving methods may be loosely thought of as ``non-stable distillation'' processes. Although SD also involves two stages -- distillation to obtain $U^\star$ and then inference on $U^\star$ -- there is no conditioning required. As is clear in our OLS model examples above, the cost of testing multiple hypotheses is not incurred until after $U^\star$ is formed. In this sense, SD is a unified testing approach, similar to knockoffs and even the humble Bonferroni method. 
	
	As Cox first noted in seminal paper on data splitting for valid post-selection inference, unified rather than two-stage inference approaches tend to enjoy more success \cite{cox_1975}. There is significant analogy here to the recent work on stable algorithms \cite{ZJ23} and data fission \cite{leiner2023data}. As thoroughly discussed in Section \ref{introduction}, these  works are fundamentally different from SD in both method and aims. Currently, SD is most naturally suited to global testing where stable algorithms and data fission are developed as variable selection methods. 
    Ideas around available data factorizations and information filtering from these approaches may be useful as SD is further developed for variable selection.

	\section{Application to Gene Testing}
	\label{sec:experiments}
	
	As described in Section \ref{sec:motivation}, identifying genes that underlie human traits and diseases via DNA sequencing datasets is a major goal of human genomics and has motivated the development of many global null testing methods \cite{wu2011rare,barnett2017generalized,liu2019acat, li2020dynamic}. Here we consider the problem of finding genes associated with a quantitative trait (a continuous outcome such as height). We leave binary traits, based on the SD procedure for logistic regression in Section \ref{sec:LR_example}, to future work. 
    
    Alongside the standard global null testing methods to which we compared SD in Section \ref{sec:power_sims}, here we also compare to a leading gene-testing method commonly used in genomics, ACAT-O \cite{liu2019acat}. ACAT-O and its generalizations, such as STAAR \cite{li2020dynamic}, have many innovations and features that are important for robust genomics analysis, including provision for close relatives and prior estimates of variant impact, which are beyond the scope of this paper. However, at its methodological core, ACAT-O relies on the Cauchy Combination Test and ANOVA-style methods. Here our aim is to show that the power advantages of SD over these methods observed in our generic regression power simulations presented in Section \ref{sec:power_sims} carry over to gene testing. Here we use realistic DNA sequencing data simulated by msprime \cite{baumdicker2022efficient,kelleher2016efficient,nelson2020accounting} to generate each design matrix $X$. For this application, we make minimal modifications to the filtered SD procedure used in Section \ref{sec:power_sims}. See Appendix \ref{sec:genomics_example} for further simulation and testing details.
	
\begin{figure}[h!t]
		\centering
		\includegraphics[width=\textwidth]{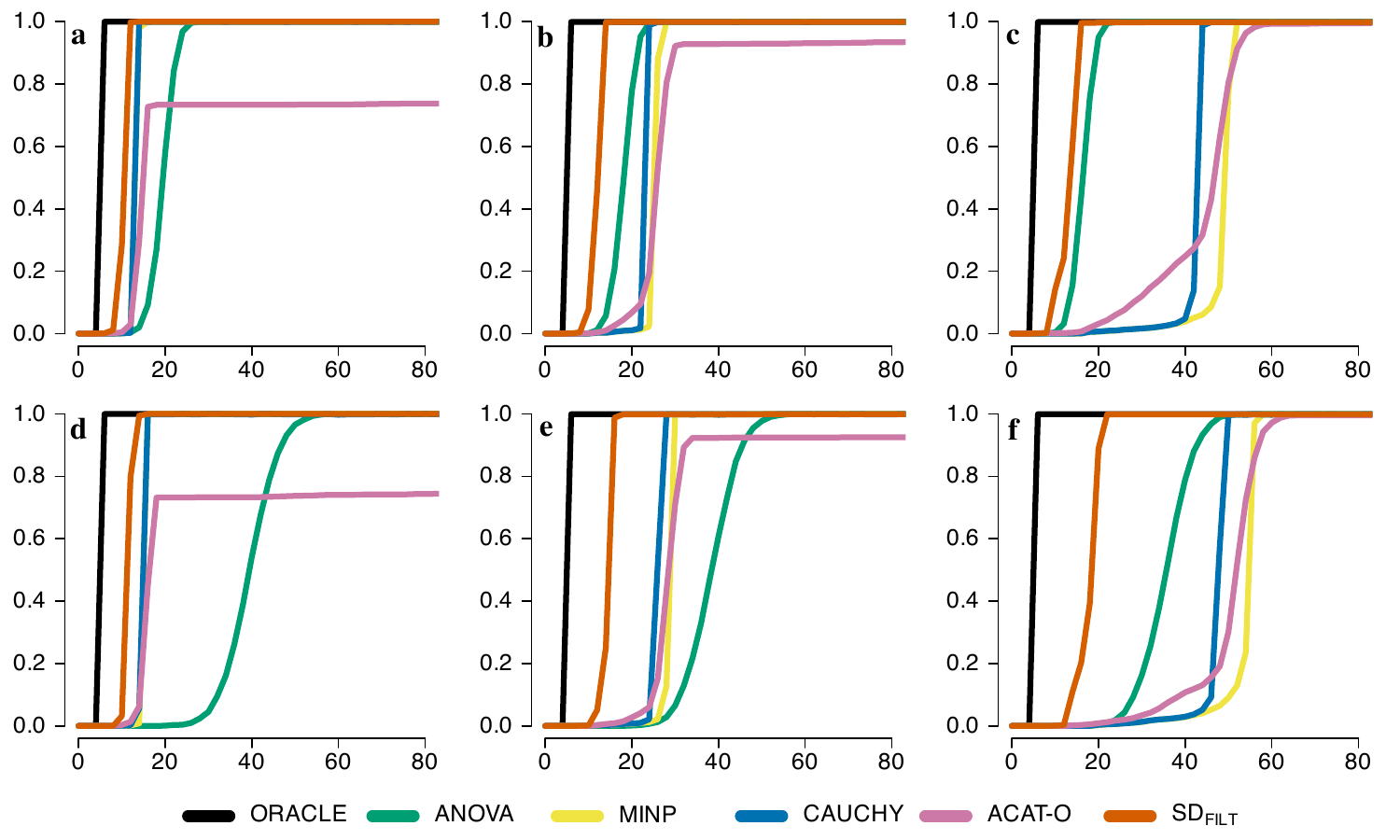}
		\caption{\small{\bf Power Curves for Gene Testing}. Each panel displays power as a function of signal strength $s$ ($s \ \approx$\ oracle $-\log_{10}$ p-value). The top row (sub-panels a,b,c) includes simulations with moderately-sized gene regions (10 kb); the bottom row, large gene regions (50 kb). The column on the left (\textbf{a} and \textbf{d}) includes simulations with 2 causal variants; the middle column (\textbf{b} and \textbf{e}), 4 causal variants; the column on the right (\textbf{c} and \textbf{f}), 8 causal variants. All simulations are based on independent datasets of 30,000 individuals generated using msprime \cite{baumdicker2022efficient,kelleher2016efficient,nelson2020accounting}. See Appendix \ref{sec:genomics_example} for details.}
		\label{fig:gene_power_curves}
	\end{figure}

	For every combination of the number of causal variants $a \in \{2,4,8\}$, gene size in $\mathrm{\{10kb, 50kb\}}$, and total signal strength $s \in \{0,2,4,6,\ldots,120\}$, we obtained 2,000 independent pairs $(Y,X)$, each corresponding to a different simulated genetic dataset. Power was calculated by estimating the probability that the p-value returned by a given testing method was less than the genome-wide gene discovery threshold $ 0.05/20,000  = 2.5 \times 10^{-6}$ from these 2,000 $(Y,X)$ pairs, yielding the power curves presented as a function of $s$ in Figure \ref{fig:gene_power_curves}. Consistent with our generic power simulations in Section \ref{sec:power_sims}, from  Figure \ref{fig:gene_power_curves} we see that $\mathrm{SD_{FILT}}$ yields more power than the alternative methods across all settings, suggesting that SD provides a more powerful basis for future gene testing approaches. 
	
	While statisticians are more comfortable with randomized procedures --- already in the 1970s Cox \cite{cox_1975} discussed the interpretation of algorithmically random $p$-values --- the inherent stochasticity of SD procedures may present an obstacle to adoption by genomics investigators or other practitioners who are rightly concerned with reproducibility: you obtain different p-values if you repeatedly run $\mathrm{SD_{FILT}}$ on the same data. A rigorous and helpful framework for practitioners interpreting these p-values is to think of the random uniform random variables generated and used during the SD procedure as \emph{part of the observed dataset}. Just as one would not collect a slightly different dataset before attempting to replicate a statistical test, when reproducing p-values produced by SD, we recommend using the same initial seed used for random number generation. However, some care around random number generation is required when using multiple-testing experiments like gene testing: different seeds should be used to initialize SD procedures at each gene in order to avoid inducing additional dependencies between gene tests. Averaging over multiple distillation runs by simply combining the resulting p-values \cite{vovk2020combining} or adapting the denoising approach proposed by Ren and Cand{\`e}s for knockoff procedures \cite{ren2023derandomizing} may help reduce the stochasticity in results produced by SD procedures.

	\bigskip
\noindent \textbf{\large Disclosure:} The authors report there are no competing interests to declare.

 \spacingset{1}
\bibliographystyle{apalike}
	\bibliography{rdistill.bib}

\clearpage

\begin{center}
\Huge Supplementary Materials
\end{center}

\appendix

\section{The R\'enyi Outlier Test} \label{sec:renyi} 
After we have used SD to decouple the parameters $\theta$, we need an outlier test like higher criticism to test $\Hg$ with power against sparse alternatives. Relatively recent work has made the null distribution of higher criticism analytically tractable for modest $p$ problems ($p \sim 10^3$) \cite{ hz19,zhang2022general}.
	Unfortunately, these calculations do not scale to very large problems ($p\sim 10^6$) and cannot reliably estimate p-values at the lower ranges accessible to standard machine precision --- on the order of $10^{-12}$ or less. The R\'enyi Outlier Test maintains comparable power against sparse alternatives but does not suffer from these constraints \cite{cox2019analysis, christ2024renyioutliertest}, which is why we will be relying upon a modified version of it to test $\Hg$ after we have used SD to orthogonalize the parameters $\theta$ rather than higher criticism.
	
	The R\'enyi Outlier Test makes use of the {\em R\'enyi transformation} $\rho : [0,1]^n \rightarrow [0,\infty]^n$. 	
    For an ordered vector $u \in [0,1]^n$, such that $u_1 \leq u_2 \leq \ldots \leq u_n$, $\rho(u)_i = i \log\left(u_{i+1} \big/ u_i\right)$ for all $i < n$ and $\rho(u)_n =  -n\log\left(u_n\right)$.
	Alfr\'ed R\'enyi pointed out that when $\rho$ is applied to a vector $U$ of ordered independent uniform random variables, the image $\rho\left(U\right)$ has entries that are independent exponential random variables \cite{AR53}. In this paper, we will make use of a modified version of the R\'enyi Outlier Test, $ \zeta_{\tau} : \left[0,1\right]^n \rightarrow \left[0,1\right]$, to map independent p-values to a single p-value. Specifically,
	\begin{equation} \label{zeta}
		\zeta_{\tau}:  u \mapsto	1-G^{-1}_{\left| \left\{u < \tau\right\}\right|+1} \bigg( \left|\left| \rho\left( \mathrm{sort}\left(\left\{ u / \tau : u < \tau \right\}\right) \right) \right|\right|_1 - \log\left(1-B_{n,\tau}\left( \left| \left\{u <\tau\right\}\right|\right)\right)\bigg),
	\end{equation}
	where $B_{n,\theta}$ is the Binomial distribution cdf with parameters $\left(n, \theta\right)$, and $G_m$ is the cdf of the Gamma distribution with rate 1 and shape parameter $m$. If $U$ is a vector of independent standard Uniform random variables, then $\zeta_{\tau}(U) \sim \mathrm{Unif}(0,1)$. The term involving $\rho$ captures information from the spacing of elements of $U$ that fall below the threshold $\tau$. The term involving the Binomial cdf feeds in information about the number of elements in $U$ that appear below the threshold $\tau$. 
	This statistic may be generalized to apply unequal weights to elements of $U$, as done in \cite{christ2024renyioutliertest}, but we will not explore this possibility here. 

    \clearpage

    \section{EFR-SD via Factorization with $L$ Iterations} \label{sec:efrsd_diagram}
	\begin{figure}[h!]
		\centering
		\begin{tikzpicture}[node distance=1cm,scale=1.5]
			\input{ro_diagram_alt.tex}
		\end{tikzpicture}
		\caption{\small \textbf{EFR-SD via Factorization with $L$ Iterations.} As in Figure \ref{F:independence}, red arrows represent the extraction step; green arrows, the filtration step; blue arrows, the reconstitution step. Here, orange arrows represent accumulating information across iterations into $U^\star$, the final step after $L$ iterations of SD.} \label{F:distdiagram}
	\end{figure}

    \clearpage

\section{Power in the case of uniformly scattered null predictors}	 \label{sec:power_appendix}
    We assume there are $K$ ``true predictors'' and $p-K$ ``null predictors'', and we perform the OLS-based SD procedure described in section \ref{sec:ols_example}.
	The null predictors $X_{\cdot j}$ (for $j\in\{1,\dots,p\}\setminus \{j_1,\dots,j_K\}$) are i.i.d.\ vectors uniformly distributed on $S^{n-1}$, the unit sphere in $\R^n$.
	The true predictors $X_{\cdot j}$ have projection $b$ onto a ``true mean'' direction $\hat\mu$, for a fixed $b\in (0,1)$; that is, $\hmt X_{\cdot j} =b$.
	The projections orthogonal to $\hat\mu$ are also uniform on the sphere, now with radius $\sqrt{1-b^2}$ in the $(n-2)$-dimensional space orthogonal to $\hat\mu$, and also independent of each other and all other predictors.
    
	Note that the SD procedure of section \ref{sec:ols_example} makes $\omega = \|Y^{(j)}\|^2$ constant.
	Let $Y^{(j)}$ be the outcome vector after $j-1$ rounds of distillation, and define $\mu_j:= \hmt Y^{(j)}/\sqrt{\omega}$.
	We assume no background covariates; that is, $q=0$.
	Let $k$ be the anticipated number of true predictors, inspiring a test at level $\alpha$ with a simple combination statistic that is $\xi=\max\sum_{l=1}^k -\log U_{j_l}$,
	where the maximum is over all choices of $k$ predictors.
	Let $\xi_\alpha$ be the critical value for the test, and define
\begin{equation} 
	\label{E:delta}
	\delta: = \frac{1-(1-b^2)^{2k_*}}{4-2b^2} - \frac43 (n\mu_1^2)^{-1} \left(\xi_\alpha + \frac{9k}{32}\right),
\end{equation}
where $k_* = \min\{k,K\}$
Note that $\delta < (1-b^2)^{k-1}$.
	
	\begin{prop}  \label{P:power}
		If 
        $\delta > 0$ and
$$
n> \max\left(\frac{12tp q_t}{\mu_1 \delta}\quad , \quad 4(tp)^2 \quad , \quad \frac{8\log t^{-1}}{b^2((1-b^2)^{k_*-1}-\delta)^2} \right),
$$
then the power of the SD procedure at level $\alpha$ is at least
\begin{equation}
	\label{E:power}
	\begin{split}
		&1-  4 \exp\left(-\frac{n\mu_1 \delta}{8} + \sqrt{n}\log\frac{\sqrt{n}\mu_1\delta}{4t}  \right)  - 12^{k_*-1}\exp\left(  - \frac{n\mu_1^2 \delta^2}{26(k_*-1)}\right)\\
         & \quad - 2k_* \exp\left( -\frac{3 n \mu_1^2}{8(1-b^2)} \left[ b^2 \left( (1-b^2)^{k_*-1} -\delta \right)^2 -
        2(n \mu_1^2)^{-1/2} (1-b^2)^{k_*-1}\sqrt{1 - 3 \log t } \right] \right) \\
        &\qquad\qquad - (2p+\ee^{tp})\exp\left(-\frac38\sqrt{n}\right) .
	\end{split}
\end{equation}
	\end{prop}
	The proof is in Appendix \ref{proof:power}. Note that $n\mu_1^2$ is in essence the absolute effect size.
	The requirement $\delta>0$ is equivalent to the expected constraint that deviations from the global null hypothesis will be detectable only if this effect size is large, on the order of the chi-square test threshold $\xi_\alpha$, which is about $k\log \frac{p}{k}-\log\alpha$; more precisely,
	\begin{lemma}
		\label{L:xicrit}
		\begin{equation}
			\label{E:xicrit}
			k\log\frac{p}{k}\le \xi_\alpha\le k\left[\log\frac{p}{k} +  \left(1+ \frac{1}{\log\frac{p}{k}} \right) \log\log\frac{p}{k} +\frac32\right] + 1-\log \alpha.
		\end{equation}
	\end{lemma}
	Proof is in Appendix \ref{proof:xicrit}.
	If there were no effect --- that is, if the global null hypothesis were true and $\hat\mu$ simply an arbitrary unit vector ---
	then $\mu_1^2$ would have Beta$(\frac12,\frac{n-1}{2})$ distribution, and $n\mu_1^2= O_{\operatorname{P}}(1)$.
	In order for there to be nontrivial power, then, this must be large.
	In the alternative model where $Y = Z + \mu$, with $Z$ a vector of i.i.d.\ standard Gaussian variables, we have $\omega = n + \|\mu\|^2 + O_{\operatorname{P}}(n^{1/2})$.
	When $\mu$ is moderately sized, so $1\ll \|\mu\| \ll n$, we have $\mu_1^2= \|\mu\|^2 \pm O_{\operatorname{P}}(\|\mu\|)$.
	
	The first term bounds the power lost in searching through potentially as many as $p$ null predictors before we get to test a true predictor.
	It will be small when the effect size $n\mu_1^2$ is large relative to $(tp)^2/n$.
    (The reason why $tp$ does not appear explicitly in this term is that we have crudely simplified the bound by imposing the constraint $tp<\sqrt{n}/2$, allowing us to express this bound in terms of $n$.)
	
	The second term is a bound on the power lost when we repeatedly test $Y$ against true predictors with correlation $b$.
	For this term to be small we need the absolute effect size to be large.
	Of course, this term may be made zero by setting $k=1$.
	That raises the question: does this power bound reveal any benefit to choosing $k>1$? that is, to combining the top $k$ p-values rather than simply looking at the smallest p-value.
	The answer is that the benefits of larger $k$ are hidden in $\delta$.
	For small $b$, $\delta$ increases with $k$, giving quantitative form to the intuition that we want $k$ to be larger when the effect size of individual predictors --- their correlation with the true mean --- is smaller, requiring us to accumulate power over multiple predictors.
	At the same time, it should be noted that we have ignored how the true predictors are oriented relative to each other.
	In a setting where we suppose the true predictors to represent different components of the true mean --- so that their correlation with each other will be small --- the bound given here would be unnecessarily crude.

    Note as well that when $k>K$ --- so, we have overestimated the number of active predictors --- we pay the full penalty for $k$ in $\xi_\alpha$, but recoup only the value of the $k_*=K$ predictors in the positive part of $\delta$.
    When we have underestimated the number of active predictors, both parts of $\delta$ correspond to the same $k$, but it is smaller than it might have been.
	
		In the setting of a linear model with fixed coefficients $\|\mu\|^2$ would be on the order of $n$, giving $n\mu_1^2$ on the order of $n^2$.
		In such a case, the power is exponentially (in $\sqrt{n}$) close to 1 as $n\to\infty$.

    The third term is a bound on the penalty for the possibility that repeated shrinking of the mean vector through repeated testing of true predictors will result in the latest ones producing p-values above the threshold.
    We crudely treat the power as 0 in this case, to avoid additional estimation problems.
    We also note that this penalty will vanish when $p$ is large, because spurious low p-values will take the place of the true predictors.
    We do not include this calculation, both because preserving power through increasing $p$ is not really the right way to think about the obstructions to power in this test, and because of the dubious validity of including the chance contributions of tests that should be null to the power.

        \clearpage
        
\section{Power Simulation Details} \label{sec:power_sim_details}

    We use a modified version of the R\'enyi Outlier Test \cite{christ2024renyioutliertest}, $ \zeta_{\tau} : \left[0,1\right]^p \rightarrow \left[0,1\right]$, to map the $10,000$ independent test statistics collected in $U^\star$ to a single p-value. See Appendix \ref{sec:renyi} for the definition of $\zeta_{\tau}$. Let $U^\star_{\tau}$ denote the $U^\star$ returned by our filtered SD procedure where $\tau$ is the quantile filtering threshold. For a desired type-1 error rate $\alpha$ and hypothesized number of active predictors $\hat{a}$, we will use threshold
	\begin{equation} \label{t}
		t_{\hat{a}} = F^{-1}_{\hat{a},p-\hat{a}+1}\left(2\alpha\right).
	\end{equation}
	To hedge against a bad choice, we will perform an omnibus test over geometrically spaced thresholds, assuming that at most $1\%$ of predictors are active: $\hat{a} \in \left\{2,4,8,16,32,64,128\right\}$. For each $\hat{a}$ we test the resulting $U^\star_{t_{\hat{a}}}$ with $\zeta_{t_{\hat{a}}}$, so that the SD filter and the final outlier test both use the same threshold $t_{\hat{a}}$. In order to cover the $a=1$ case, we also incorporate the Bonferroni corrected p-value $U^\dagger$ calculated among the marginal test statistics for each predictor. 
	Taking the minimum and the corresponding Bonferroni correction, we arrive at
	\begin{equation} \label{bf_sd_stat}
		8 \min \left( U^\dagger, \zeta_{t_{2}}\left(U^\star_ {t_{2}}\right), \zeta_{t_{4}}\left(U^\star_ {t_{4}}\right), \zeta_{t_{8}}\left(U^\star_ {t_{8}} \right),\ldots,\zeta_{t_{128}}\left(U^\star_ {t_{128}}\right) \right)
	\end{equation}
	as our p-value based on filtered SD. For comparison, we take $\zeta_{1}(U^\star_1)$ as our p-value for na\"ive (non-filtered) SD. For each replicate of the filtered SD procedure and the na\"ive SD procedure, we sampled a permutation from $\mathcal{S}_p$ uniformly at random to dictate the order in which to distill the simulated predictors.
	
	We fix each column of $A \in \mathbb{R}^{n \times 3}$ by simulating three vectors: an intercept, a vector of standard Gaussian variables, and a vector of Rademacher random variables, all independent. We simulate $X$ so that the columns follow a block covariance structure based on clusters of size 10. More explicitly, the entries of $X \in \mathbb{R}^{n \times p}$ are drawn from centered Gaussian variables so that the covariance structure follows $\mathbb{E}(X^\top X) = I_{\tfrac{p}{10} \times \tfrac{p}{10}} \bigotimes B$ where $B \in \mathbb{R}^{10 \times 10}$, $B_{ij} = \begin{cases}
		r & i\neq j \\
		1 & i = j
	\end{cases}$
	for some positive $r$. In our simulations, we specify three levels of dependence within the blocks of predictors in terms of the proportion of variance shared, $r^2 \in \left\{0.2, 0.5, 0.8\right\}$.  We set the nuisance parameters from \eqref{model} $\sigma = 2$ and $\alpha = \mathbf{1}$. In each simulation, we select a desired number of active predictors $a \in \left\{4,16\right\}$ uniformly at random from $[p]$. 
	Given this active set $\mathcal{A}$ we select $\beta$, while aiming to distribute the observed effects across active predictors as evenly as possible. To accomplish this, consider the $QR$-decomposition $QR = \wt{X}_\mathcal{A}$, which we define as $P_A^\bot X_\mathcal{A}$ with length-normalized columns. Given $\sigma$, the sufficient statistic for the oracle model is $ \left|\left| Q^\top Y\right|\right|_2^2$ with expected value $\left|\left| R \beta \right|\right|_2^2 $. For a desired underlying ``signal strength'' $s$, we solve $R \beta =\sqrt{F^{-1}_{\chi^2_a}\left(1-10^{-s}\right)} \ \mathbf{1}$ to make the magnitude of each entry of $\beta$ as similar as possible. Then, we simulate $Y = A\alpha + \sigma\left(\wt{X}_\mathcal{A} \beta  + P_{X_\mathcal{A}}^\bot \epsilon\right)$ where $\epsilon \sim N(0,I_n)$. This ensures that the observed $\hat{\beta} = R^{-1}Q^\top Y = \sigma \beta$ is stable across simulations and that the $-\log_{10}$ p-value that one would obtain by testing the resulting $Y$ with an oracle ANOVA model that ``knows'' the active predictors will be approximately $s$ in every simulation.

    \clearpage

    \section{$n<p$ Power Simulation Results} \label{sec:dense_small_n_sims}
    
\begin{figure}[!ht]
		\centering
		\includegraphics[width=0.95\linewidth]{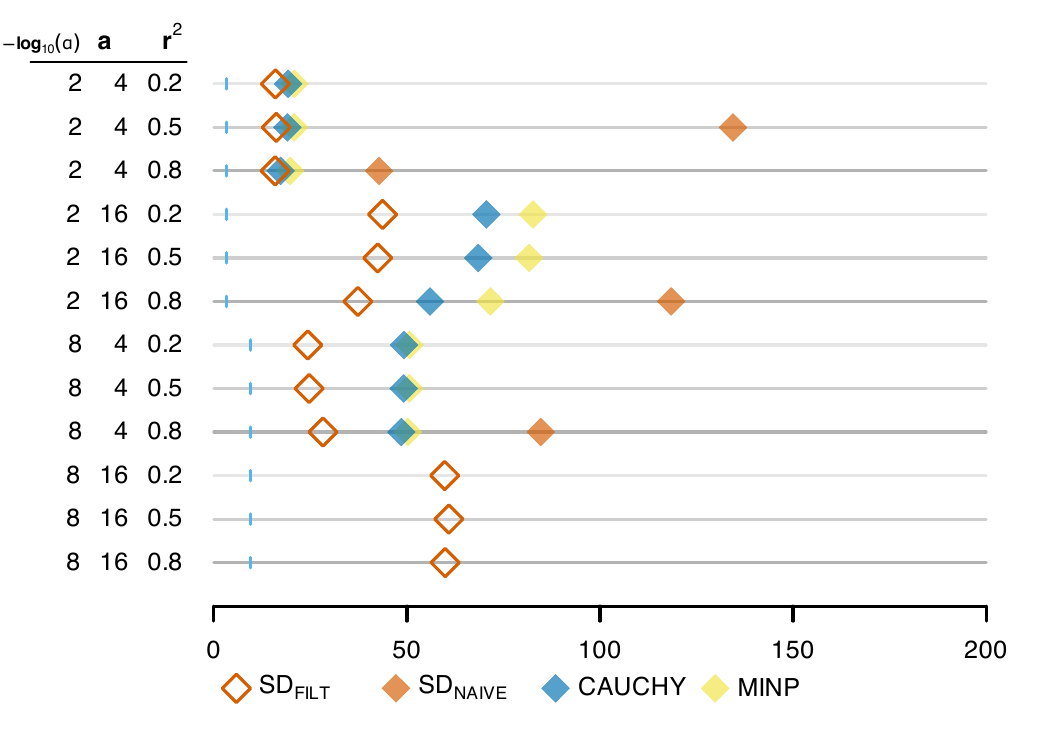}
		\caption{\small \textbf{Power in the $n<p$ case}. Dotplot of signal strength ($s \ \approx$\ oracle $-\log_{10}$ p-value) required to achieve 80\% power under various simulation conditions. Here, $p = 10,000$ and $n=1,000$. The columns of the key on the left list the size of the test ($10^{-2}$ or $10^{-8}$), the number of active predictors ($a$), and the squared correlation between predictors within each block/cluster $r^2$ respectively. The light blue tick marks on each line indicate the effect size at which the oracle model achieves 80\% power. The absence of a diamond on any line indicates that it is beyond the range of the plot (signal strength $>$ 200).}
		\label{fig:dense_small_n_sims}
	\end{figure}

    Here we present our power simulations for the case where $n < p$. ANOVA is omitted since it is undefined for the $n<p$ case. In Figure \ref{fig:dense_small_n_sims} we see a dramatic deterioration in the power of naïve SD, and that filtering rescues the SD procedure. Given the low dimensionality of each $n$-vector relative to $p$, it makes sense that modifying the projection of $Y$ onto predictors early in the distillation procedure would have a relatively large impact on the projection of $Y$ onto predictors encountered later in the distillation procedure. When the size of the test is $10^{-8}$ and there are 16 active predictors only filtered SD attains significant power anywhere near the range of oracle signal strength.

    \clearpage
    
    \section{Computational Scalability} \label{sec:scalability}
\begin{figure}[!ht]
		\begin{minipage}[c]{0.5\linewidth}
			\includegraphics[width=\linewidth]{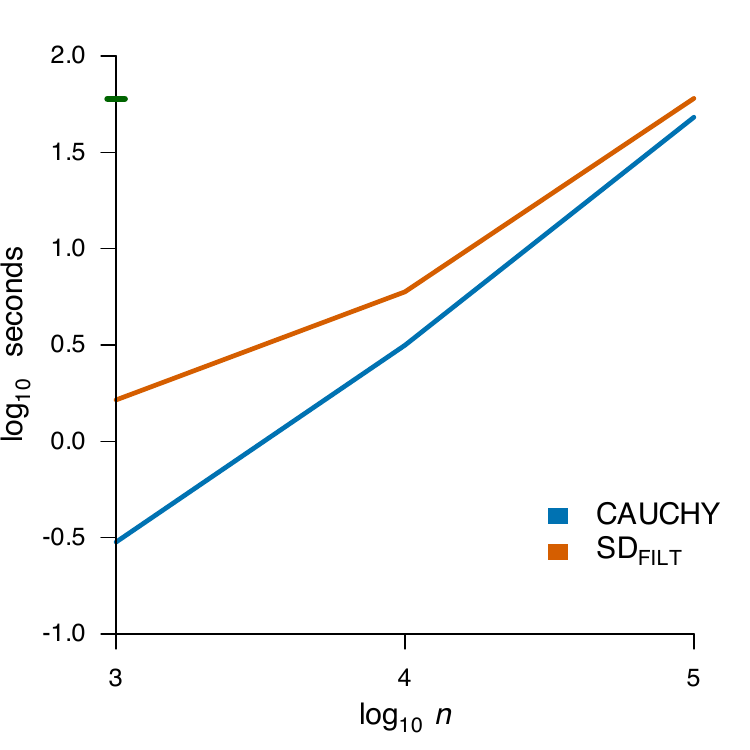}
		\end{minipage}\hfill
		\begin{minipage}[c]{0.5\linewidth}
			\caption{\small \textbf{Benchmarking.} Median run time, in terms of $\log_{10}$(seconds), as a function of sample size, in terms of $\log_{10}(n)$. The green notch on the y-axis marks one minute.
			} \label{fig:benchmark}
		\end{minipage}
	\end{figure}

	Here we briefly compare the scalability of our SD procedure to the Cauchy Combination Test. The run time of the Cauchy method is effectively equal to the time required to compute the marginal p-values for each predictor; combining them with the Cauchy CDF is nearly instantaneous. 
	Here we benchmark our \texttt{R}-based implementations of both approaches. 
	In these benchmarking simulations we fix $p = 10^4$ and $q = 3$, simulating $A$ as above. We consider three problem sizes: $n \in \left\{10^3, 10^4, 10^5\right\}$. All benchmarks were performed on a machine with 32 Intel(R) Xeon(R) Gold 6242 CPUs, each at 2.8GHz. \texttt{R} was compiled against OpenBLAS, which was configured to use all available cores. The median run times, each based on 50 independent replicates, are shown in a log-log plot in Figure \ref{fig:benchmark}. We see that our \texttt{R}-implementation of the Cauchy method scales linearly in $n$ (the slope is 1), as expected since most of the compute is handled by the highly optimized BLAS. While there is significant overhead in our \texttt{R}-implementation of the filtered SD procedure, as can be seen from the relatively long run time when $n=10^3$, we see that the relative cost of this overhead diminish as $n$ increases and the BLAS operations begin to dominate the run time. For large $n$, we expect both the Cauchy method (based on $p$ BLAS-parallelized projections) and our filtered SD implementation to scale as $\mathcal{O}(np)$. For large $n$, we not only see similar scaling for the two algorithms in $n$, but also that the run time of SD is very close to that of the Cauchy method. For $n=10^5$, we observe a Cauchy median run time of 48.1s and an SD median run time of 60.3s.

    \clearpage
    \section{Availability of a General Decomposition}\label{sec:general_decomposition}
Suppose we have the general random variables $Y:\Omega \to \Y$ and $U=U(Y)$ where $U:\Omega \to \U$.
	Following our general assumption that $\Y$ and $\U$ are Polish spaces, these define countably generated $\sigma$-algebras $\eu{Y}$ and $\eu{U}$, and $(\Y,\eu{Y},\P)$ is a regular conditional probability space \cite[Theorem 5.1.9]{rD10}.
	Hence, by \cite{dR79} there exists an {\em independent complement} $\eu{V}$ to $\eu{U}$; that is, a sub-$\sigma$-algebra of $\eu{Y}$ independent of $\eu{U}$, such that $\eu{V}$ and $\eu{U}$ jointly generate $\eu{Y}$ modulo null sets.
	If $\eu{V}$ is countably generated --- for example, if $\Omega$ itself is countable --- then we can use it to define a random variable $V$ on some space $\V$ that is independent to and ``complementary'' to $U$, in the sense that $(V,U)$ can be used to reconstruct $Y$ almost surely.
	Actually constructing this $V$ is difficult, in general, and while the EFR construction does not require that $V$ be calculated explicitly, these complications invite a more intuitive generalization based on coupled random variables. If we have a single random variable $U$ that has been split off from $Y^{(\ell)}$, the function of an independent complement is essentially to couple $Y^{(\ell)}$ to a new $Y$ that will be generated from a newly generated value $\wt{U}$. In the main text, we show how this coupling can be made explicit. 

    \clearpage
\section{Maximal Coupling Equations for Logistic Regression}\label{sec:logistic_regression_coupling_equations}
  Here we consider a random walk $S_k := \sum_{i=1}^k Y_i X_{ij}$ starting from $i=1$ and progressing to $i=n$. We are free to randomly order the samples along the walk; preliminary experiments suggest that starting the random walk with samples where $\hat{p}_i$ is close to 0 or 1 improves efficiency. Define $q_k(s):= \P\bigl( \sum_{i=k+1}^n Y_i X_{ij} =s \bigr).$
	Then, having observed a partial sum $S_{i-1}=s$, we may compute probabilities for the next step $Y_i=0$ or 1 by
	\begin{equation} \label{coupling_prob}
		\P \bigl( Y_i = 1 \, \bigm| \, S_{i-1}=s, \, S_n=r \bigr) = \frac{\hat p_i q_i(r-s - X_{ij})}{q_{i-1}(r-s)}.
	\end{equation}
	 Given a new $\wt{W} = \wt{Y}^\top X_j = \wt{r}$, we can couple the steps of a new random walk $\wt{S}_k := \sum_{i=1}^k \wt{Y}_i X_{ij}$ to the steps of the original random walk $\left(S_k\right)_{i=1}^n$ by sampling from the following joint mass function. The coupling is ``maximal'' in the sense that it maximizes the probability of the same direction for each step $i$.
     
     Let $\kappa\left(s,r\right)$ be a function that maps to \eqref{coupling_prob}. We achieve a maximal coupling between our new $\wt{Y}$ and the original $Y$ via
	\begin{align*}
		\P \bigl( Y_i = \wt{Y}_i = 1 \, \bigm| \, S_{i-1}=s, \, S_n=r, \, \wt S_{i-1}=\tilde s, \, \wt S_n=\tilde r \bigr) &= \kappa\left(s,r\right) \wedge \kappa\left(\tilde{s},\tilde{r}\right),\\
		\P \bigl( Y_i = \wt{Y}_i = 0 \, \bigm| \, S_{i-1}=s, \, S_n=r, \, \wt S_{i-1}=\tilde s, \, \wt S_n=\tilde r \bigr) &= 1-  \kappa\left(s,r\right) \vee \kappa\left(\tilde{s},\tilde{r}\right),\\
		\P \bigl( Y_i = 1,\,  \wt{Y}_i = 0 \, \bigm| \, S_{i-1}=s, \, S_n=r, \, \wt S_{i-1}=\tilde s, \, \wt S_n=\tilde r \bigr) &= \bigl(\kappa\left(s,r\right) - \kappa\left(\tilde{s},\tilde{r}\right) \bigr)_+,\\
		\P \bigl( Y_i = 0,\,  \wt{Y}_i = 1 \, \bigm| \, S_{i-1}=s, \, S_n=r, \, \wt S_{i-1}=\tilde s, \, \wt S_n=\tilde r \bigr) &= \bigl(\kappa\left(\tilde{s},\tilde{r}\right) - \kappa\left(s,r\right) \bigr)_+.
	\end{align*}

    \clearpage
	\section{Consolidating SD iterations}\label{sec:dimensionality}
	\subsection{Introduction to consolidation} \label{sec:partitionintro}
	Extracting more information relevant to $\Hg$ with each move of the SD Markov chain, and thereby consolidating the number of SD iterations, can improve the power of SD by reducing the amount of noise introduced over the course of the distillation process. Increasing the dimensionality of each $U^{(l)}$ has an intuitive motivation in the context of the OLS model. If $X$ followed an orthogonal design and there were no background covariates $A$, then all of the $\hat{\beta}$ would be independent, all of the predictors could be tested in parallel with a single iteration of SD, and no noise from $U'$ would be injected into the estimated $\hat{\beta}$. Here, we revisit the OLS model where we have grouped the predictors into $L \leq p$ layers. We start by showing how the bijective factorization function $\psi$ for the OLS model we presented in Proposition \ref{prop_1_var}  can be generalized to handle multiple predictors in parallel. We then provide a formal description of and heuristics for partitioning predictors into layers.
    
	\subsection{Distilling the OLS model with multiple predictors per layer} 
	
	To distill multiple predictors per layer under the OLS model \eqref{model}, we use the orthogonal design case as inspiration. By ``orthogonal design,'' we do not mean simply that the columns of $X$ are orthogonal, but additionally that no row of $X$ has more than one non-zero entry. The analyst can group (partition) the predictors arbitrarily, but to maximize power the predictors should be partitioned with a view toward minimizing the shared variance among predictors in the same layer.
	
	Consider first a version of the OLS model \eqref{model} with three simplifying restrictions: $\alpha = 0$, $\sigma = 1$, and the columns of $X$ can be partitioned into layers such that the predictors within each layer follow an orthogonal design. 
	In this special case, it would be simple to generalize from distilling one predictor per layer to multiple predictors per layer since the estimators $\hat{\beta}_j$ within each layer would be independent. 
	We could then ``assign'' each sample $i$ to the unique predictor $j$ such that $X_{ij} \neq 0$. 
	
	Relaxing these assumptions requires that we now formalize this notion of assigning samples to predictors using assignment vectors $\xi$. 
	In the definition below we account for the general case with background covariates $A$ by performing a separate projection for each predictor in a layer. 
	The processes of partitioning predictors, assigning samples to predictors, and projecting predictors to account for $A$ interact jointly with the goal of maximizing power.
	For this reason, we present these projections and assignment vectors conjointly with our definition of a partitioning below.
	
	A {\em partitioning} $\pi_A(X)$ is a collection, one for each layer $l\in\{1,\dots,L\}$, of $\xi^{(l)}: [n] \to [p]$, understood as assignments of each subject to a single predictor:
	\begin{equation*}
		\pi_A(X) = \left(\xi^{(1)}, \xi^{(2)}, \ldots, \xi^{(L)} \right).
	\end{equation*}
	We say that predictor $j$ {\em is in} (or {\em belongs to}) layer $l$ if $j$ is in the image $\xi^{(l)}([n])$. 
	We use $S_l := \left\{j \in [p] : j \in \xi^{(l)}[n] \right\}$ to denote the set of predictors in layer $l$ and call $p_l := | S_l |$ the {\em size} of layer $l$. 
	The set of samples assigned to predictor $j$ (in the unique layer that predictor $j$ belongs to) will be called $G_j$:
	\begin{equation*}
		G_j := \left\{i \in [n] : \xi^{(\{l : j \in S_l\})}_{i} = j \right\}.
	\end{equation*}
	
	\begin{definition} \label{def:partitioning}
		$\pi_A(X)$ is a {\em partitioning} of $X \in \mathbb{R}^{n \times p}$ with respect to $A \in \mathbb{R}^{n \times q}$ under the OLS model with $L$ layers if
		\begin{enumerate}
			\item the images $\xi^{(l)}([n])$ comprise a partition of $\{1,\dots,p\}$; that is, $\bigcup_l \xi^{(l)}([n]) = \{1,\ldots,p\}$ and $\xi^{(l)}([n]) \cap \xi^{(k)}([n]) = \emptyset \ \forall \ l\neq k$;
			\item $q < \left|G_j\right| \ \forall \ j \in \left\{1,\ldots,p\right\}$;
			\item $\wt{X}_j \neq 0 \ \forall \ j \in \left\{1,\ldots,p\right\}$
		\end{enumerate}
		where 
		\begin{equation*}
			\wt{X}_j = \ddot{X}_{\cdot j} / \bigl\|\ddot{X}_{\cdot j} \bigr\| \quad \mathit{and} \quad  
			\ddot{X}_{ij}=
			\begin{cases}
				\left[ P^\bot_{A_{G_j \cdot }} X_{G_jj}\right]_i \quad i \in G_j \\
				0 \quad \mathrm{otherwise}
			\end{cases}.
		\end{equation*} 
	\end{definition}
	This construction ensures that the implied $W_j = Y^\top \wt{X}_j$ are non-degenerate. 
	The more general definition of $\wt{X}_j$ introduced here is consistent with our prior definition of $\wt{X}_j$ for the case of one predictor per layer. 
	Since samples can only be assigned to at most one predictor, $\wt{X}_{\cdot S_l}$ follows an orthogonal design while ensuring $\wt{X}^\top_{\cdot S_l} A  = 0$ for each $l$. 
	This ``generalized orthogonal design''  guarantees the independence of the implied $W_j$ within a layer despite the presence of background covariates.
	
	The assumption that $P^\bot_A X_{\cdot j} \neq 0$ for all $j = 1,\ldots,p$ that we introduced with the OLS model \eqref{model} guarantees that we can always find a valid $\pi_A(X)$; for example, we could simply place one predictor in each layer and, for each layer, trivially assign every sample to the sole predictor in that layer. The trade-off in terms of power going from this one predictor per layer partitioning to a more consolidated partitioning depends on the balance between the noise removed in the SD procedure by the consolidation and the power lost by shoehorning predictors into an orthogonal design within each layer. 
	While it is difficult to quantify the former, the latter can be expressed in terms of the proportion of projected variance assigned to each predictor. 
	We calculate this below in Section \ref{sec:partitioning}, and provide further practical guidance on partitioning predictors, and demonstrate that the trade-off can strongly favour consolidation in the case where $X$ follows an orthogonal design. 
	
	While our partitioning definition above guarantees independence among the implied $W_j$ in the presence of background covariates, defining a valid SD procedure requires that we also account for the unknown scale $\sigma$ of the $W_j$.
	A naive approach would be to studentize each $W_j$ separately, using only the subset of samples assigned to predictor $j$. 
	This would be equivalent to performing separate $t$ tests for each predictor in the layer. 
	It is clear, though, that this procedure would lose substantial power when a predictor is sparse, with only a handful of samples assigned to it, hence leaving very few degrees of freedom to estimate $\sigma$.
	We would like to share degrees of freedom for estimating $\sigma$ as far as possible across predictors within a layer, while maintaining independence. 
	To do this, we adapt the following representation of the standard i.i.d. Gaussian vector, taking our inspiration from the stick-breaking representation of the Dirichlet Process. Here we let $\mathcal{G}_n = [0,1]^{n-1} \times \{-1,0,1\}^{n} \times \R_{+}$.
	
	\begin{lemma}
		\label{lem:stick_breaking}
		For an $n$-dimensional random vector $W$ let $\omega := \left|\left|W\right|\right|^2_2$ and $B_j := W_j^2 \ / \ \sum\limits_{i = j}^n W_j^2$ for $j = 1,\ldots,n-1$. Let $\mu_n : \mathbb{R}^n \rightarrow \mathcal{G}_n$ be the bijective map \begin{equation}
			\mu_n(W) = \left(B_1,\ldots B_{n-1}, \sgn\left(W_1\right),\ldots,\sgn\left(W_n\right), \omega \right) = \left(B, \sgn\left(W\right), \omega\right).
		\end{equation} If $W \sim N(0,I)$, then all elements of $\mu_n(W)$ are mutually independent where $\sgn\left(W_j\right) \sim \mathrm{Rademacher}$, $B_j \sim \mathrm{Beta}\left(\frac{1}{2},\frac{n-j}{2}\right)$, and $\omega \sim \chi^2_{n}$.
	\end{lemma}
	From this lemma, we quickly arrive at a bijective factorization that we can use for distillation under the OLS model, once we have specified  a permutation that prioritizes predictors within the layer $\lambda_l : \mathcal{S}_{p_l} \to [p]^{p_l}$.\footnote{To connect these beta statistics to the more classic T statistic from \eqref{classic_t}, note that $B_1$ is a simple monotonic transformation of $T_1^2$: $B_{1} = \frac{(n-q-1) T_{\lambda_l(1)}^2}{1+(n-q-1) T_{\lambda_l(1)}^2}$. Effectively $n-q-1$ degrees of freedom are used to estimate the unknown $\sigma$ when estimating $\beta_{\lambda_l(1)}$. However, only $n-q-p_l$ degrees of freedom are used to estimate $\sigma$ when estimating $\beta_{\lambda_{l(p_l)}}$. While this inhomogeneity may appear problematic at first, note that only $n-q-p_l$ degrees of freedom are used to estimate $\sigma$ in standard ANOVA. The difference in power across predictors within a layer is negligible as long as no layer has too many predictors. For this reason and general power considerations, as a general rule of thumb, we recommend selecting a partitioning $\rho_A(X)$ where $ n-q-\max\limits_lp_l \geq 50$.} Recall that $F_{a,b}$ is the CDF of the Beta distribution with shape parameters $(a,b)$ and mean $\tfrac{a}{a+b}$. Note $1 - F_{a,b} = F_{b,a}$.

	\begin{prop}  \label{P:covar3}
		Let $\lambda_l : \mathcal{S}_{p_l} \rightarrow [p]^{p_l}$. Let orthonormal matrices $\wt{X}_{\cdot \lambda_l} \in \mathbb{R}^{n \times p_l}$ and $V \in \mathbb{R}^{n \times \left(n - p_l - q\right)}$ form an orthnormal basis $\left[\wt{X}_{\cdot \lambda_l},V\right]$  for $H = \left\{ v \in \mathbb{R}^{n} : A^\top v = 0\right\}$. Let $\eta: \mathcal{G}_{n-q} \rightarrow \mathcal{G}_{n-q}$ where $\left[\eta(x)\right]_j = F_{\frac{n-j}{2},\frac{1}{2}}(x)$ for $j < n$ and $x_j$ otherwise. Define $\psi^\star: \R^n \to \R^q \times \mathcal{G}_{n-q}$ by
		
		\begin{align}
			\psi^\star(Y) &= \left( P_A Y, \ \eta \ \circ\  \mu_{n-q}\left(\left[\wt{X}_{\cdot\lambda_l},V\right]^\top Y \right) \right) \\ 
			\nonumber &= \left( P_A Y,\ \eta \ \circ\  \mu_{n-q}\left(W \right)\right) \\
			\nonumber &= \left( P_A Y,\ U_1,\ldots, U_{n-1}, \sgn\left(W_1\right),\ldots,\sgn\left(W_n\right), \omega \right)  \\ \nonumber &= \left( P_A Y,\ U, \sgn\left(W\right), \omega\right).
		\end{align}
		If $Y \sim N\left(A\alpha, \sigma I_n\right)$, then all elements of $\psi^\star(Y)$ are mutually independent where $U_j \sim \mathrm{Unif}(0,1)$, $\sgn\left(W_j\right) \sim \mathrm{Rademacher}$, and $\omega/\sigma \sim\chi^2_{n}$.
	\end{prop}
	
	Applying this factorization $\psi^\star$ in practice does not require explicit construction of $V$. Let $\nu = \omega - \sum_{k=1}^{p_l} W_{\lambda_l(k)}^2$. Under $H_g$, $\nu/\sigma \sim \chi^2_{n-q-p_l}$ and is independent of the $W_j$ for $j = 1,\ldots,p_l$. It is sufficient to calculate these $W_j$ and summarize $\omega$ and the remaining $W_{p_l + 1},\ldots, W_{n}$ via $\nu$.
	
	To invert this procedure given $\wt{U}$, we simply use the corresponding beta quantile functions to obtain $\wt{B}_1,\ldots,\wt{B}_j$ and set $\wt{W}_{j}^2 = \omega \wt{B}_j \prod\limits_{i=1}^{j-1} \left(1 - \wt{B}_i\right)$. From here, we calculate $\wt{\nu} = \omega - \sum\limits_{j=1}^{p_l} \wt{W}_{j}^2$. Then we update $Y$ with what is only a slight extension of \eqref{eq:simple_inversion}:
	\begin{equation}
		\wt{Y} = P_A Y +  \sqrt{\tfrac{\wt{\nu}}{\nu}} \left(P^\bot_A Y -  \wt{X}^\top_{\cdot \lambda_l} W\right) + \wt{X}^\top_{\cdot \lambda_l} \wt{W}.
	\end{equation} 
	As above, this procedure maintains $\wt{Y}^\top\wt{Y}  = \omega$.
	
	
	\subsection{Partitioning}\label{sec:partitioning}
	
	While our distillation procedure is valid for any partitioning $\pi_A(X)$, we aim to select a $\pi_A(X)$ that maximizes power to reject $H_g$ when distillation is iterated over the layers. This involves a trade-off in minimizing the influence of two opposing sources of noise.
	
	On one side, we have the familiar noise that the statistician cannot control: the noise inherent to the data generating process that is endogenous to the observed data. Minimizing the impact of this noise involves extracting as much information relevant to $\Hg$ in each layer as possible. In the notation of Definition \ref{def:partitioning}, for the OLS model, this means assigning as many samples as possible to each predictor so that $\left\| \ddot{X}_{ij} \right\|_2$ is not much smaller than $\bigl\| P_A^\bot X_{j} \bigr\|_2$. 
	In other words, we maximize the effective sample size for performing inference about each $\beta_j$. 
	If we were concerned only to reduce this source of noise, we would run SD with the trivial partitioning of one predictor per layer.
	
	On the other side we have the exogenous noise that we inject into our inference during the SD process. Minimizing the impact of this noise involves extracting the information most relevant to $\Hg$ in early layers and minimizing the total number of layers, $L$. For the OLS model, this translates to a) prioritising predictors that are suspected of having non-negligible effects in one of the early layers and b) packing the predictors into as few layers as possible. 
	Overall, this helps to ensure that each $\beta_j$  is estimated with a $Y^{(l)}$ that is as similar to the initially observed $Y^{(0)}$ as possible. 
	Were we concerned only to reduce this source of noise, we would  pack all of the predictors into a single layer, assigning relatively few, or sometimes even no, samples to some predictors. 
	
	We can calibrate the trade-off between minimizing endogenous and exogenous noise in our choice of $\pi_A(X)$. Whenever possible, a practioner should start by examining the structure of $X$ to see if there are any sets of orthogonal columns or other patterns that could make it straightforward to construct layers such that the predictors within each layer are nearly orthogonal. However, in the absence of any obvious patterns, we need an automated procedure for selecting $\pi_A(X)$ for the OLS model, assuming the use of a quantile filter at each iteration. 
	Our overall strategy may be applied to other models, and aims simply to minimize exogenous noise subject to an upper bound on increasing the endogenous noise.
	
	In order to measure the impact of endogenous noise, note that the power of SD to accumulate evidence against $\Hg$ when using a quantile filter with threshold $t$ is dependent on $\p{U_j < t}$ for all $j : \beta_j \neq 0$. 
	For the OLS model, this probability varies as a function of the true underlying $\beta_j$ and the $\pi_A(X)$. To more easily demonstrate the influence of $\pi_A(X)$ on $\p{U_j < t}$, let us momentarily suppose that we have placed all of our predictors in a single layer and that there is only one active predictor $j$; that is, one predictor such that $\beta_j \neq 0$. This allows us to drop the subscript $j$ and write $Y$ in terms of a single column $X$ corresponding to the active predictor. In this simplified case, applying the projection $P^\bot_A$ to both sides of \ref{model}, we have
	
	\begin{equation}
		P^\bot_A Y = \beta \bigl\|P^\bot_A X \bigr\| \frac{P^\bot_A X }{\|P^\bot_A X \|} + P^\bot_A \epsilon.
	\end{equation}
	
	Now consider setting a $\pi_A(X)$ that assigns all samples to the active predictor and ignores all of the other predictors. This would maximize the effective sample size for and minimize the endogenous noise involved in estimating $\beta$. After renormalizing the predictor $P^\bot_A X$ as above, we would be targeting an implied effect size $\beta_{[n]} := \beta \bigl\|P^\bot_A X \bigr\|$. Define the function $f_t$ by
	\begin{equation}
		f_t(\beta):= \p{\left(Z +\beta \right)^2 > F^{-1}_{\chi^2_1}\left( 1-t\right)  }  \quad \textrm{where} \quad Z\sim N(0,1) 
	\end{equation}
	so that $\p{U < t} = f_t\left(\beta_{[n]}\right)$. 
	If under $\pi_A(X)$, only the subset $\xi$ of samples are assigned to predictor $X$, we have 
	\begin{equation}
		P^\bot_{A_{\xi\cdot}} Y_{\xi\cdot} = \beta \bigl\|P^\bot_{A_{\xi\cdot}} X_{\xi\cdot} \bigr\| \frac{P^\bot_{A_{\xi\cdot}} X_{\xi\cdot} }{\|P^\bot_{A_{\xi\cdot}} X_{\xi\cdot} \|} +P^\bot_{A_{\xi\cdot}} \epsilon_\xi
	\end{equation}
	with an implied effect size $ \beta_\xi = \beta \bigl\|P^\bot_{A_{\xi\cdot}} X_{\xi\cdot} \bigr\| $. So if we have a true effect $\beta$ that yields some initial detection probability $p_0 = f_t\left(\beta_{[n]}\right)$, we can guarantee a lower bound on the detection probability, $p_1$, after assigning a subset of samples $\xi \subseteq [n]$ to a given predictor by requiring
	
	\begin{equation}
		\frac{\bigl\|P^\bot_{A_{\xi\cdot}} X_{\xi\cdot} \bigr\|}{\|P^\bot_{A} X \|} \geq  \frac{f^{-1}_t\left(p_1\right)}{f^{-1}_t\left(p_0\right)}.
	\end{equation}
	
	Returning to the general context where we have a partitioning with multiple predictors and multiple predictors may have non-zero effects, in the notation of Definition \ref{def:partitioning}, this amounts to enforcing a lower bound on 
	$\| \ddot{X}_{ij} \| \big/ \|P^\bot_{A} X_j \|$ while building $\rho_A(X)$. Such constraints can be enforced via various greedy or integer programming procedures while seeking to minimize the total number of layers. 
    
    It is possible to actively allocate samples to predictors within a layer that appear ``interesting'' based on putative test statistics within our SD framework, as is done in the high-dimensional knockoff filter \cite{barber2019}. We leave development of procedures using such an allocation scheme to a future paper.
	
\clearpage

\section{Avoiding degeneracy under the logistic regression model} \label{sec:avoiding_degeneracy_for_logistic_regression}
As discussed in section 6 of the main text, our treatment of SD for logistic regression rapidly loses power if the design matrix entries are too various.
There need to be multiple alternative random paths leading to the same weighted sum, so that we can select one that is biased to be similar to the original data.
In the extreme case, where $X$ contains arbitrary real entries, each possible sum will be obtainable in exactly one way, and the ``coupling'' will be singular.
The resampled outcome data will be completely decoupled from the original data, and power will be essentially 0.

If the $X$ entries are not all integers, but are all similarly sized, one possible solution would be to treat the entries $X_{ij}$ as being themselves subject to random noise.
That is, instead of testing the statistic $W_j=\sum X_{ij} Y_i$, we test $\wt W_j := \sum X_{ij}(1+\sigma \xi_{ij}) Y_i$,
where $(\xi_{ij})$ are i.i.d.\ standard Gaussian random variables.
This will have variance $(1+\sigma^2)$ times the usual variance of the score test statistic.
The residual sum $S_k=\sum_{i=k}^n X_{ij}(1+\sigma \xi_{ij}) Y_i$ has a continuous distribution supported on all of $\R$, so we may compute conditional probabilities of $Y_k=1$ conditioned on $\{S_k=U-s\}$, and so define a coupling as above.
The power to identify effective predictors will presumably depend on a favourable choice of $\sigma$: If $\sigma$ is too small, the information in the data $Y$ will be rapidly subsumed in noise; if too large, the information in the design matrix $X$ will evaporate.

It is not clear whether this Gaussianized coupling can be computed efficiently.
Certainly there does not seem to be an obvious way of adapting the FFT algorithm from the Poisson-binomial case.

\clearpage

\section{Smallest-$k$ Filtering} \label{sec:smallest_k_filtering}
This appendix builds up increasingly complex filtering procedures that target exchanging the smallest-$k$ order statistics between $U$ and $U'$. We start with ``exchanging the smallest-$k$ ratios''. For the $l^{th}$ layer with $p_l$ predictors, as above, let $T$ be the order statistic of $U$ so that $T_1 \leq T_2 \leq \ldots \leq T_{p_l}$. Say in a given layer, we can safely assume that the number of non-null hypotheses is less than some $k$. Then, we can set $\varphi\left(T\right)= \left(P,Q\right)$ where
\begin{equation}\label{E:k1}
	\begin{array}{ll}
		P_j=T_{j} & \mathrm{ \ for \ } j = k+1,\ldots,p, \\
		Q_j= T_j / T_{k+1} &\mathrm{ \ for \ } j = 1,\ldots,k,
	\end{array}
\end{equation}
and the $R_1,\ldots,R_m$ are null. Since $R_1,\ldots,R_m$ are null, $\S$ is not needed or defined in this example.
Note that $Q$ here is distributed as the order statistics of $k-1$ independent standard uniform random variables.
This definition of $\varphi$ ensures the required independence between $P$ and $Q$; however, since it always exchanges the tail of $T$ with the tail of $T'$, it can introduce a fair amount of noise into the SD process over many iterations.
It also can lose substantial power if $k$ is chosen too small.

We can make the procedure more robust against mis-specification of $k$ by moving $T_k$ from $P$ to $Q$ so that it is always exchanged during filtration.
This means, in effect, that the smallest-$k$ uniform random variables rather than the smallest $k$ ratios are exchanged. Let $g(x,c) = (x-c) / (1-c)$ where $c \in [0,1]$. We set $\varphi(T) = (P,Q)$ where
\begin{equation}\label{E:k2}
	\begin{array}{ll}
		P_j=g\left(T_{j}, T_k \right)& \mathrm{ \ for \ } j = k+1,\ldots,p_l, \\
		Q_j= T_j &\mathrm{ \ for \ } j = 1,\ldots,k,
	\end{array}
\end{equation}
and the $R_1,\ldots,R_m$ are null. Again, this definition of $\varphi$ ensures the required independence between $P$ and $Q$. When we invert $\varphi$ in this case, we see that all of the original order statistics $T_{k+1},\ldots,T_{p_i}$ are shifted and rescaled according to the simulated $k^{th}$ order statistic. In the OLS context, this makes $\wt{W}_j \neq W_j$ for all predictors $j$ in layer $l$. This additional noise allows us to extract the absolute values of the $k$ smallest $U'$, which would all still be outliers in $U$ even if we under-specified $k$, giving us the desired robustness to $k$. However, if we do not under-specify $k$ and $p_i$ is large, 
$\mathrm{Var}( T_k - \wt{T}_k)$ will be small.
This makes the difference between $\wt{W}_j$ and $W_j$ small with high probability, so the noise we need to inject in order to exchange $T_k$ during filtration can actually be smaller than it might be supposed. Like \eqref{E:k1}, since \eqref{E:k2} essentially always exchanges the tail of $T$ with the tail of $T'$, it can introduce a fair amount of noise into the distillation process over many iterations.

By introducing  non-trivial $R_1,\ldots,R_m$, \eqref{E:k1} and \eqref{E:k2} can be generalised to reduce the information exchanged during filtration. We start by generalizing \eqref{E:k1}. Let $G_j = \left(T_j /T_{j+1}\right)^j$ for $j \in \left\{1,\ldots,k\right\}$, and fix some $c_j\in [0,1]$. By properties of the R\'enyi transformation \cite{AR53} discussed in Appendix \ref{sec:renyi}, under $\Hg$, $G_j \overset{\mathrm{iid}}{\sim} \mathrm{Unif}\left(0,1\right)$. We can set $\varphi\left(T\right)= \left(P,Q,R_1,\ldots,R_m\right)$ where
\begin{equation}\label{E:k3}
	\begin{array}{ll}
		P_j=T_{j} & \mathrm{ \ for \ } j = k+1,\ldots,p_i, \\
		Q_j= \mathbf{1}\left\{G_j \leq c_1 \right\}  &\mathrm{ \ for \ } j = 1,\ldots,k, \\
		R_j = \left(G_j / c_1\right)^{Q_j} g\left(G_j,c_1\right)^{1-Q_j} &\mathrm{ \ for \ } j = 1,\ldots,k.
	\end{array}
\end{equation}
Here, $R$ has $m=k$ components. With this definition of $\varphi$ we can simply set our selection function to be $\S(Q) = Q$. With this selection function, if $Q_j = 0$, then, upon inversion of $\varphi$, $G_j$ is returned unchanged. In other words, this choice of $\varphi$ and $\S$ ensures that ratios between consecutive $T_j$ are only exchanged if they are more extreme than some quantile $c_1$.

The underlying motivation for introducing this $c_1$: if a given gap is not mildly significant, it does not provide evidence against $H_g$, so exchanging it would only introduce unnecessary noise into $Y$. Note that \eqref{E:k3} reduces to \eqref{E:k1} when $c_1 = 1$.
If $c_1$ is set to 0.1, then under the null, on average, only outlying 10\% of ratios are exchanged between $T$ and $T'$. We can reduce this exchange further by lowering $c_1$  to 0.01 so that, under the null, only the most outlying 1\% of ratios are exchanged.
However, if $c$ is set too small then the exchange of information between $T$ and $T'$ can be so limited that we lose power.
In all of our simulations, we will set $c_1=0.05$ as a default. If $c_1=0.05$ and $k = 8$, then under $H_g$, the probability that inverting \eqref{E:k3} returns $\wt{Y} = Y$ is $(1-0.05)^8 \approx 0.66$.
Even if $k=16$, the probability only drops to $(1-0.05)^{16} \approx 0.44$.
In other words, with this smallest-$k$ filter, SD can often screen a large number of predictors $p_l$ for ``free'' -- without introducing any noise into $Y$. Here, even when noise is introduced, it only modifies the most extreme (smallest) $U$. Recall, we see a similar result from filtering with a quantile filter.

By adding one additional parameter, $c_2$, we can further generalise \eqref{E:k3} so that it includes \eqref{E:k2} as a special case. Let $G'_j = F_{j,p-j+1}\left(T_j\right)$ (recall $F_{a,b}$ is the CDF of the beta distribtion). For our general ``top-k'' filter we have
\begin{equation}\label{E:k4}
	\begin{array}{ll}
		P_j= g\left(T_{j}, T_k \right) & \mathrm{ \ for \ } j = k+1,\ldots,p_i, \\
		Q_j= \mathbf{1}\left\{G_j \leq c_1 \right\}  &\mathrm{ \ for \ } j = 1,\ldots,k-1, \\
		Q_k= \mathbf{1}\left\{G'_k\leq c_2 \right\}  & \\
		R_j = \left(G_j / c_1\right)^{Q_j} g\left(G_j,c_1\right)^{1-Q_j} &\mathrm{ \ for \ } j = 1,\ldots,k-1,\\
		R_k = \left(G'_k / c_2\right)^{Q_k} g\left(G'_k,c_2\right)^{1-Q_k}. & \
	\end{array}
\end{equation}
Again, we set $\S(Q) = Q$ and note that \eqref{E:k4} reduces to \eqref{E:k2} when $c_1=1$ and $c_2 = 1$. There are many more sophisticated examples of $\varphi$ and $\S$ available: the tails of $U$ and $U'$ can be exchanged in blocks rather than element-by-element as in \eqref{E:k3} and \eqref{E:k4} by making use of Lukacs's proportion-sum independence theorem \cite{lukacs}. The development of different pairs of filtration and selection functions is left for future work.

\clearpage
\section{Genomics Simulations} \label{sec:genomics_example}

\subsection{Haplotype Simulation}

We simulated 2,000 genomic datasets using \texttt{msprime} \cite{kelleher2016efficient,nelson2020accounting,baumdicker2022efficient}. Each dataset consisted of a 1Mb chromosome observed for 30k human samples. In order to model the diversity of arising genomic datasets, 10k samples in each dataset were drawn from each of three 1000 Genomes Populations -- Yoruba (YRI), Han Chinese (CHB), and Central European (CEU). This was done based on the Demography tutorial provided in the \texttt{msprime} documentation \cite{msprimetutorial}, which itself was based on the population parameters related those three populations presented in \cite{gutenkunst2009inferring}. Explicitly, 10k samples in each dataset were drawn from each of three 1000 Genomes Populations -- Yoruba (YRI), Han Chinese (CHB), and Central European (CEU). We only made one modification to the demographic model specified in \cite{msprimetutorial}: we use the Discrete Time Wright-Fisher model for the first 100 generations into the past before reverting back to the classic Hudson model as proposed in \cite{nelson2020accounting}.

Each 1Mb region of simulated haplotypes was generated using a population-specific human recombination map estimated by \texttt{pyrho} \cite{spence2019inference}. Explicitly, in each simulation, we randomly selected one of our three 1000 Genomes Populations (YRI, CHB, or CEU) and 1Mb segment from genome (excluding heterochromatic regions), then we loaded the recombination map estimated for that population and region by \texttt{pyrho}.

In real analyses, variants that are only observed on one chromosome in a given sequencing dataset are often considered likely to be introduced by sequencing error and removed from downstream analyses. Accordingly, we remove any variants present in only one copy (singletons) from our simulated datasets.

For each simulated dataset we extracted a region within in the middle of the chromosome to be our gene of interest. We considered two gene sizes: 10 kb or 50 kb. As is the standard default when using ACAT-O and other gene-testing methods, we removed all common variants (with minor allele frequency $\geq 0.01$) from the gene, leaving only the rare variants (with minor allele frequency $< 0.01$) for testing. Due to the inherent stochasticity in simulating mutations, the total number of rare variants in the gene varies across datasets. Let $p^{(i)}_{10}$ be the number of rare variants in the 10kb gene segment of simulated dataset $i \in \{1,2,\ldots,2000\}$. Define $p^{(i)}_{50}$ analogously for each 50 kb gene segment. Our aim is to test $\Hg$ under our OLS model \eqref{model} where the matrix of predictors in simulation $i$, $X^{(i)} \in \{0,1,2\}^{n \times p^{(i)}}$, is the genotype matrix of rare variants within the gene region. Across our 2000 simulations, $p^{(i)}_{10}$ ranged from 100 to 178 with a mean of 129.9; $p^{(i)}_{50}$ ranged from 554 to 745 with a mean of 647.4. 

\subsection{Phenotype Simulation}
To simulate our outcome vector $Y$, we selected causal variants uniformly at random from among the rare variants encoded by the columns of $X$ under the restriction that the column of $X$ (the genotype vector) corresponding to each causal variant must be distinct. We considered the case of $a = $ 2, 4, or 8 active predictors (causal variants). We simulated our matrix of background covariates $A$ just as we did in  Section \ref{sec:power_sims}, except that we added two covariates corresponding to population indicators in order to allow for different intercepts for samples from YRI, CHB, and CEU. Outside of that adaptation of $A$, we assigned effects to our causal variants and simulated $Y$ exactly as in Section  \ref{sec:power_sims}. As further described in that section, this procedure guaranteed the contribution to the total signal strength $s$ as defined in Section  \ref{sec:power_sims} of each causal variant would be roughly equal in every simulation.

\subsection{Testing Details}
\begin{figure}[h!]
	\centering
	\includegraphics[width=0.7\textwidth]{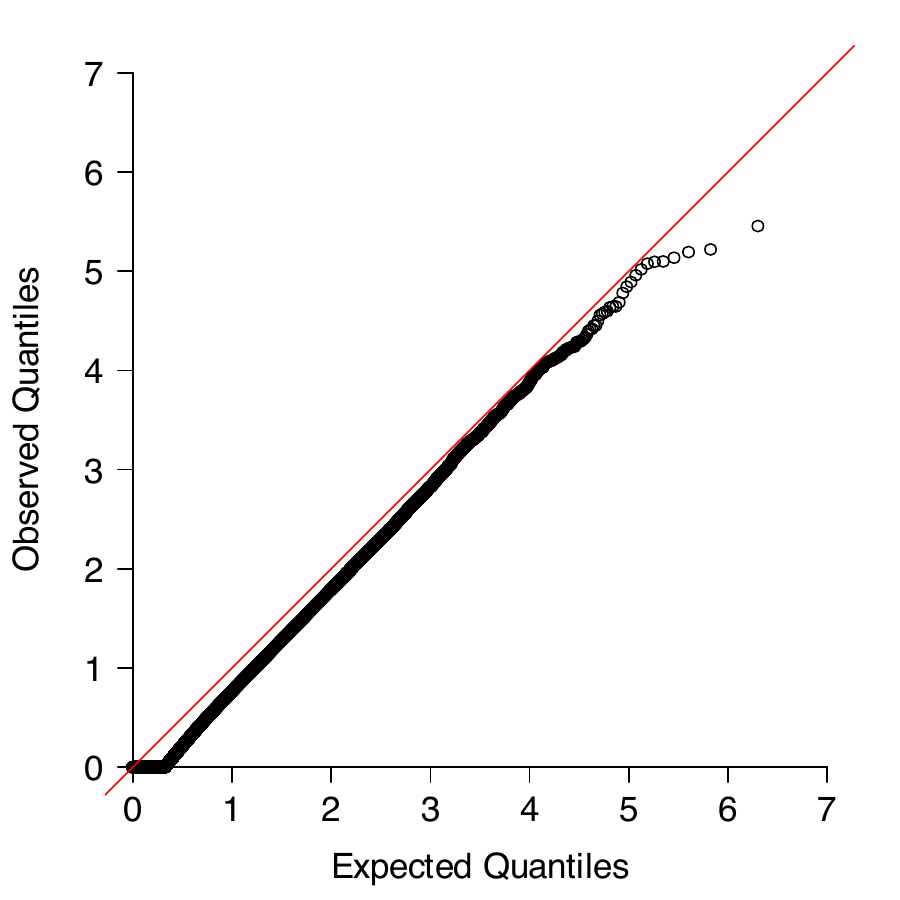}
	\caption{{\bf Confirming SD Calibration for Gene Testing Simulations}. QQ plot of the $-\log_{10}$ p-values returned by $\mathrm{SD_{FILT}}$ across 1 million null simulations using our gene simulation and testing procedure.}
	\label{fig:gene_null_sims}
\end{figure}

For each simulation, we tested $\Hg$ using the oracle ANOVA model (ORACLE), the Cauchy Combination Test (CAUCHY), and the Bonferroni-adjusted minimum marginal p-value (MINP) as done in Section \ref{sec:power_sims}. Here we also ran ACAT-O with all default settings. We compared these methods to the simple filtered SD procedure ($\mathrm{SD_{FILT}}$) presented in Section \ref{sec:power_sims} with two context-specific customizations. First, we used a multiple testing threshold $\alpha = 0.05 / 20000 = 2.5 \times 10^{-6}$ corresponding to the number of genes tested in a typical gene-focused association study to calculate each threshold $t$ $\eqref{t}$ rather than $10^{-2}$ or $10^{-8}$. Second, we also thought it unrealistic that any gene would be expected to have $64$ or $128$ causal variants, so we only considered cases where the estimated number of active predictors was $\hat{a} \in \{1,2,4,8,16,32\}$ when calculating thresholds via $\eqref{t}$. Accordingly, since here we only considered 6 rather than 8 values of $\hat{a}$, we multipled by 6 rather than by 8 when combining SD results via \eqref{bf_sd_stat}. We confirmed that our implementation of this procedure yielded a calibrated p-values via 1 million null simulations. See Figure \ref{fig:gene_null_sims}.

\clearpage 
\hypertarget{proofs}{%
\section{Proofs}\label{proofs}}

\raggedbottom
\subsection{Proof of Lemma \ref{L:Uindependence}}
{\ }\\
{\em Statement of Lemma}: \quad
If $(Y^{(l)},U^{(l)})_{l=1}^L$ is a SD process, then $ U^{(1)}, U^{(2)}, \ldots, U^{(L)}$ are mutually independent.\\

\noindent {\em Proof}: \quad We show by induction that the collection of random variables $(U^{(l)})_{l=k}^L$ are mutually independent for each $k\in [L]$. The statement is trivial for our base case $k=L$. Assume it is true for a given $k>1$ (Fact A). Consider the $k-1$ case. Since a SD process is a HMM by definition, the family $(U^{(l)})_{l=k}^L$ is d-separated from $U^{(k-1)}$ by $Y^{(k)}$, hence the family $(U^{(l)})_{l=k}^L$ is independent of $U^{(k-1)}$ conditioned on $Y^{(k)}$ (Fact B) by Theorem 3.3 of \cite{KF09}. 
Also by the definition of a SD process, $U^{(k-1)}  \indep Y^{(k)}$  (Fact C). Let $f_{k-1},f_{k},f_{k+1},\ldots, f_{L}:[0,1] \to \R$ be any measurable bounded functions.

\begin{align*}
	\E\left[\prod\limits_{j=k-1}^L f_{j}(U^{(j)}) \right]  &= \E\left[ \E\left[ \left.  f_{k-1}(U^{(k-1)}) \prod\limits_{j=k}^L f_{j}(U^{(j)}) \right| Y^{(k)} \right]  \right] \\
	& = \E\left[ \E\left[ \left.  f_{k-1}(U^{(k-1)})  \right| Y^{(k)} \right]  \E\left[ \left. \prod\limits_{j=k}^L f_{j}(U^{(j)}) \right| Y^{(k)} \right]  \right]  \quad \textrm{by Fact B}\\
	&= \E\left[ \E\left[ f_{k-1}(U^{(k-1)})  \right]  \E\left[ \left. \prod\limits_{j=k}^L f_{j}(U^{(j)}) \right| Y^{(k)} \right]  \right] \quad \textrm{by Fact C} \\
	&= \E\left[ f_{k-1}(U^{(k-1)})  \right]  \E\left[ \E\left[ \left. \prod\limits_{j=k}^L f_{j}(U^{(j)}) \right| Y^{(k)} \right]  \right]  \\
	&= \E\left[ f_{k-1}(U^{(k-1)})  \right]  \E\left[\prod\limits_{j=k}^L f_{j}(U^{(j)}) \right]  \\ 
	&= \E\left[ f_{k-1}(U^{(k-1)})  \right]   \prod\limits_{j=k}^L \E\left[ f_{j}(U^{(j)})  \right] \quad \textrm{by Fact A}.
\end{align*}
Hence $\{U^{(k-1)},\dots, U^{(L)}\}$ are mutually independent, completing the induction. $\square$ \newline

\subsection{Proof of Theorem \ref{T:independentEFR}}
{\ }\\
{\em Statement of Theorem}: \quad
The EFR-SD procedure satisfies the conditions for generating a stable distillation if the following hold:
\begin{itemize}
	\item  The two components $\psi(Y)=(V,U)$ are independent;
	\item $U$ and $\wt{U}$ have the same distribution;
	\item The two components $\wt{U}$ and $U^{(l)}$ are independent.
\end{itemize}

{\ }\\
\noindent {\em Proof:} \quad
The assumption of independent noise implies the necessary conditional independence for $Y^{(l)}\mapsto (U^{(l)},Y^{(l+1)})$ to continue the HMM.

$\wt{U}$ is a deterministic function of $U$, $U'$, and $\omega$, which are jointly independent of $V$, hence $\wt{U}$ is independent of $V$. By the first two conditions of the lemma then $(V,\wt{U}) \overset{d}{=} (V,U)$, hence $Y^{(l+1)}= \psi^{-1}(V,\wt{U}) \overset{d}{=}  \psi^{-1}(V,U) =Y^{(l)}$.

Finally, $V$ is independent of the pair $(\wt{U},U^{(l)})$, implying by the last condition that all three variables are mutually independent, so that $U^{(l)}$ is independent of $(V,\wt{U})$.
Hence $U^{(l)} \indep Y^{(l+1)}$. \qed

\subsection{Proof of Theorem \ref{T:generalEFR}}
{\ }\\
{\em Statement of Theorem}: \quad
The general EFR-SD procedure satisfies the conditions for generating a stable distillation if the following hold:
\begin{itemize}
	\item  The joint distribution $\pi_{u,\tilde{u}}(\diff y, \diff \tilde y)$ has marginal distributions $\pi_u(\diff y)$ and $\pi_{\tilde u}(\diff \tilde y)$ for every choice of $u$ and $\tilde u$;
	\item The joint distribution $\f(\diff u, \diff \tilde u, \diff u^{(l)})$ has marginal distribution $q$ in each component, and makes $\wt{U}$ and $U^{(l)}$ independent.
\end{itemize}

{\ }\\
\noindent {\em Proof:} \quad
Let $\F_{l}^<$ be the $\sigma$-algebra generated by $(Y^{(1)},\dots,Y^{(l)})$, $\F_\ell$ the $\sigma$-algebra generated by $Y^{(l)}$, and let $U$ be the random variable $U(Y^{(l)})$.
Then by construction $U^{(l)}$ has distribution conditioned on $\F^<_{l}$ given by $\f'_U$, which is measurable with respect to $\F_l$, hence $U^{(l)} \indep (Y^{(1)},\dots, Y^{(l-1)})$ conditioned on $Y^{(l)}$.
Similarly, $Y^{(l+1)}$ has distribution conditioned on $\F^<_{l}$ given by $\f_U(\diff \tilde u) \pi_{\tilde u, Y^{(l)}}(\diff y^{(l+1)})$, which is also measurable with respect to $\F_l$, hence also $Y^{(l+1)} \indep (Y^{(1)},\dots, Y^{(l-1)})$ conditioned on $Y^{(l)}$.
This proves by induction the HMM property.

Now let $f:\U\to \R$ and $g:\Y\to\R$ be bounded measurable test functions.
Then
\begin{align*}
	\E\bigl[f(U^{(l)}) g(  Y^{(l+1)})\bigr] & = \int f(u^{(l)}) g(\tilde y) \f( \diff u,\diff \tilde u, \diff u^{(l)})  \pi_{u,\tilde u} (\diff y, \diff\tilde y)\\
	&= \int f(u^{(l)}) g(\tilde y) \f( \diff u,\diff \tilde u, \diff u^{(l)}) \pi_{\tilde u} (\diff \tilde y),
\end{align*}
after carrying out the integration in $y$, as the coupling $\pi_{u,\tilde u}(\diff y,\diff \tilde y)$ has the same marginal $\pi_{\tilde u}(\diff \tilde y)$ for each $u$.
As $u$ now no longer appears in the integrand we may integrate over $u$, reducing the joint distribution to the paired marginal of $(\tilde{u},u^{(l)})$, which is $q(\diff\tilde{u})q(\diff u^{(l)})$.
Thus
\begin{align*}
		\E\bigl[f(U^{(l)}) g(  Y^{(l+1)})\bigr] &=  \int f(u^{(l)}) g(\tilde y) q( \diff \tilde u) q( \diff u^{(l)}) \pi_{\tilde u} (\diff \tilde y) \\
	&= 	\E\bigl[f(U^{(l)})\bigr]\cdot \E\bigl[ g(  Y^{(l+1)})\bigr] ,
\end{align*}
showing that $U^{(l)}$ and $Y^{(l+1)}$ are independent. \qed

\subsection{Proof of Proposition \ref{prop_1_var}}
{\ }\\
{\em Statement of Proposition}: \quad \\
The map $\psi_j: \mathfrak{N} \to  \R^q  \times (0,1] \times \{-1,0,+1\} \times \R_+\times \mathfrak{S}$ given by \\
$\psi_j: Y \mapsto \left(P_A Y, \ U_j, \ \sgn\left(T_j\right), \ \omega , \ P^\bot_{AX_{\cdot j}} Y / \left|\left|P^\bot_{AX_{\cdot j}} Y\right|\right|_2 \right)$ is invertible. 
When $Y \sim N(A\alpha,\sigma I_n)$, the five components of $\psi_j(Y)$ are mutually independent.\\

\noindent {\em Proof}: \quad
It is sufficient to show that $\psi_j$ is the composition of bijective functions. 
The mapping $f : \R^n \to \R^q \times \R \times \R^{n-q-1} $ given by 
$y \mapsto \bigl(P_A y, \ P_{\wt{X}_j} y, \ P_{A,\wt{X}_j}^\top y \bigr)$ is bijective by the Rank-Nullity theorem, since it projects $\R^n$ onto disjoint subsets whose total dimension is $n$.
Restricted to $\mathfrak{N}$, $f$ is a bijective map onto $\R^q \times \R \times \left\{\R^{n-q-1} \setminus 0\right\}$.

The map $g : \R^{n-q-1} \setminus \{0 \} \to \R_+ \times \mathfrak{S}$ given by 
$ x  \mapsto \left(\left|\left|x\right|\right|_2^2, \frac{x}{\left|\left|x\right|\right|_2} \right)$ is bijective. 
Composing $g$ and $f$, where we apply $g$ to the last component of $f(Y)$, we have a map $Y\mapsto \bigl(P_A Y, \ P_{\wt{X}_j} Y, \ \nu , \ P^\bot_{AX_{\cdot j}} Y / \bigl\|P^\bot_{AX_{\cdot j}} Y\bigr\|_2 \bigr)$. 
Note, in our notation introduced in the main text, $W_j = P_{\wt{X}_j} Y$ and $\omega = \nu + W_j^2$. 

We now define the bijective map $h : \R \times \R_+ \to (0,1] \times \{-1,0,1\} \times \R_+$ by $$
h(x_1,x_2) = \Bigl( \frac{x_1^2}{x_1^2 + x_2}, \ \sgn(x_1), \ x_1 + x_2 \Bigr).
$$
This gives us $h(W_j,\nu) = \left(B,\sgn\left(W_j\right),\omega\right)$ where $B = W_j^2 / \omega$. 
Note that $\sgn\left(W_j\right) = \sgn\left(T_j\right)$. 
Observing that $T_j^2 = (n-q-1) \frac{B}{1-B}$ is invertible and $U_j$ is an invertible function of $T_j^2$ proves that $\psi_j$ is invertible.

We now consider the distribution of $\psi_j(Y)$ when $Y \sim A\alpha + \sigma Z,$ where $Z \sim N\left(0,I_n\right)$. 
Since orthogonal projections of $Z$ are independent and

$$\left(P_A Y, \ P_{\wt{X}_j} Y, \ P_{A,\wt{X}_j}^\top Y \right) \sim \left(A\alpha + P_A Z, \ P_{\wt{X}_j} Z, \ P_{A,\wt{X}_j}^\top Z \right)$$

\noindent we have that $\left(P_A Y, \ P_{\wt{X}_j} Y, \ P_{A,\wt{X}_j}^\top Y \right)$ are mutually independent. 
Since $P_{A,\wt{X}_j}^\top Y$ is a standard multivariate normal in the null space of $\left(A,\wt{X}_j\right)$, it is spherically symmetric. 
Hence, we have independence between length, $\nu = \bigl\|P_{A,\wt{X}_j}^\top Y\bigr\|_2^2$, and direction, $P_{A,\wt{X}_j}^\top Y \big/ \bigl\| P_{A,\wt{X}_j}^\top Y\bigr\|_2$.
It follows that the components of $g\circ f$ are independent.
We now apply the function $h$, to the second component, which we denote by $W_j$.
Since $W_j \sim N(0,1)$, $\sgn\left(W_j\right)$ is independent of $W_j^2$, so $\sgn\left(W_j\right)$ is independent of $B$ and $\omega$. 
Since $B \sim \mathrm{Beta}\left(\frac{1}{2},\frac{n-q-1}{2}\right)$,  we have $B \indep \omega$ from Lukacs's Proportion-Sum Independence Theorem \cite{lukacs}. 

This establishes that the map from $Y$ to $$\bigl(P_A Y, \ B,\ \sgn\left(T_j\right), \ \omega, \ P^\bot_{AX_{\cdot j}} Y / \bigl\|P^\bot_{AX_{\cdot j}} Y\bigr\|_2 \bigr)$$ has independent components. 
The final mapping acts only on the component $B$, thus preserves independence.
\qed

\subsection{Proof of Lemma \ref{L:xicrit}}

{\em Statement of Lemma}:

	We suppose that we expect $k$ true predictors, hence have decided to test at level $\alpha$ with a simple combination statistic that is $\xi=\max\sum_{l=1}^k -\log U_{j_l}$,
	where the maximum is over all choices of $k$ predictors.
	Let $\xi_\alpha$ be the critical value for the test.

		\begin{equation}
			k\log\frac{p}{k}\le \xi_\alpha\le k\left[\log\frac{p}{k} +  \left(1+ \frac{1}{\log\frac{p}{k}} \right) \log\log\frac{p}{k} +\frac32\right] + 1-\log \alpha.
		\end{equation}

\begin{proof}
	Under the global null hypothesis $U_1,\dots,U_p$ are independent uniformly distributed on $[0,1]$, hence $-\log U_1,\dots,-\log U_p$ are independent exponentially distributed with parameter 1.
	Thus $\xi$ is the sum of the top $k$ order statistics of independent exponentials.
	Using the R\'enyi representation for these order statistics we may write
	$$
	\xi=  \sum_{i=0}^{p-k} \frac{k}{p-i} \tau_i + \sum_{i=p-k+1}^{p-1} \tau_i ,
	$$
	where $\tau_0,\dots,\tau_{p-1}$ are i.i.d.\ exponential with parameter 1.
	The moment generating function is
	$$
	M_\xi(s) = (1-s)^{-k}\prod_{i=0}^{p-k-1} \frac{p-i}{p-i-k s} \text{ for } 0<s<1.
	$$
	An application of Chernoff's inequality shows that the tail probability above $\xi_\alpha$ is less than $\alpha$.
\end{proof}

\subsection{Proof of Lemma \ref{L:xicrit}}\label{proof:xicrit}
{\ }\\
{\em Statement of Proposition}:
{\ }\\
		\begin{equation}
			k\log\frac{p}{k}\le \xi_\alpha\le k\left[\log\frac{p}{k} +  \left(1+ \frac{1}{\log\frac{p}{k}} \right) \log\log\frac{p}{k} +\frac32\right] + 1-\log \alpha.
		\end{equation}
	
	\begin{proof}
		Under the global null hypothesis $U_1,\dots,U_p$ are independent uniformly distributed on $[0,1]$, hence $-\log U_1,\dots,-\log U_p$ are independent exponentially distributed with parameter 1.
		Thus $\xi$ is the sum of the top $k$ order statistics of independent exponentials.
		Using the R\'enyi representation for these order statistics we may write
		$$
		\xi=  \sum_{i=0}^{p-k} \frac{k}{p-i} \tau_i + \sum_{i=p-k+1}^{p-1} \tau_i ,
		$$
		where $\tau_0,\dots,\tau_{p-1}$ are i.i.d.\ exponential with parameter 1.
		The moment generating function is
		$$
		M_\xi(s) = (1-s)^{-k}\prod_{i=0}^{p-k-1} \frac{p-i}{p-i-k s} \text{ for } 0<s<1.
		$$
		An application of Chernoff's inequality shows that the tail probability above $\xi_\alpha$ is less than $\alpha$.
	\end{proof}

\subsection{Proof of Proposition \ref{P:power}}\label{proof:power}
{\ }\\
{\em Statement of Proposition}:
{\ }\\

		If 
        $\delta > 0$ and
$$
n> \max\left(\frac{12tp q_t}{\mu_1 \delta}\quad , \quad 4(tp)^2 \quad , \quad \frac{8\log t^{-1}}{b^2((1-b^2)^{k_*-1}-\delta)^2} \right),
$$
then the power of the SD procedure at level $\alpha$ is at least
\begin{equation*}
	\begin{split}
		&1-  4 \exp\left(-\frac{n\mu_1 \delta}{8} + \sqrt{n}\log\frac{\sqrt{n}\mu_1\delta}{4t}  \right)  - 12^{k_*-1}\exp\left(  - \frac{n\mu_1^2 \delta^2}{26(k_*-1)}\right)\\
         & \quad - 2k_* \exp\left( -\frac{3 n \mu_1^2}{8(1-b^2)} \left[ b^2 \left( (1-b^2)^{k_*-1} -\delta \right)^2 -
        2(n \mu_1^2)^{-1/2} (1-b^2)^{k_*-1}\sqrt{1 - 3 \log t } \right] \right) \\
        &\qquad\qquad - (2p+\ee^{tp})\exp\left(-\frac38\sqrt{n}\right) .
	\end{split}
\end{equation*}

\newcommand{\bmu}{\beta^\mu}
\newcommand{\bp}{\beta^\perp}
\begin{proof}
	Define
	\begin{align*}
		\bmu_j &= \sqrt{n}\hmt X_j,\\
		\bp_j &= \sqrt{n-1}\frac{(Y_{j}^\top - \mu_{j} \hmt \sqrt\omega ) X_j}{\sqrt{\omega(1 - \mu_{j}^2)}},\\
		\beta_j &= \sqrt{n}\frac{Y_{j}^\top X_j}{\sqrt\omega}  = \mu_{j} \bmu_j + \sqrt{1 - \mu_{j}^2} \bp_{j}.
	\end{align*}
	The procedure described in section \ref{sec:ols_example} works with the test statistic ${T_j^2=\beta_j^2/(n-1-\beta_j^2)}$, which is F-distributed with $(1,n-1)$ degrees of freedom under the global null hypothesis.
	Let $q_t/(n-1-q_t)$ be the $1-t$ quantile of this F distribution, so that $\P(\beta_j^2 > q_t) =t$.
	As $T_j^2$ is a monotonic function of $\beta_j^2/n$, so we may simplify our notation slightly by treating $\beta_j^2$ as the test statistic.
	Each component $\beta^{2}/n$ and $\beta^{\mu2}/n$ has beta distribution with parameters $(\frac12,\frac{n-1}{2})$, so $q_t$ may also be understood as a quantile of this beta distribution.
	The orthogonal component $\beta^{\perp2}/(n-1)$, as it is conditioned to lie in an $(n-1)$-dimensional subspace, has beta distribution with parameters $(\frac12,\frac{n-2}{2})$.
	
	When $X_j$ is a null predictor our assumptions imply that $T_j^2$ is independent of $Y^{(j)}$ and still has the same F distribution, and the components $\beta^2/n$ have the $\operatorname{Beta}(\frac12,\frac{n-1}{2})$ distribution.
	The probability of exceeding the filtering threshold $q_t$ is thus exactly the nominal probability $t$, and the cdf transformation that is supposed to produce substitute p-values does in fact yield values uniform on $[0,1]$.
	
	In the following calculations we drop the $j$ from the notation of all $\beta$ and $X$ terms, for simplicity.
	We generate an independent $\beta'$ with $\beta^{'2}/n$ having distribution Beta$(\frac12,\frac{n-1}{2})$.
	There are now three things that could happen in the transition from $Y^{(j)}$ to $Y^{(j+1)}$:
	\begin{enumerate}
		\item With probability $t$ the test statistic $\beta^2$ exceeds the threshold $q_t$. $U_j$ is returned as the beta cdf of $\beta^2/n$, and the square projection of $Y^{(j)}$ in the direction of $X$ is replaced by $\omega$ times an independent Beta random variable with the same sign as $\beta$, and the orthogonal projection rescaled to preserve the length of $Y$.
		\item If $\beta^2 < q_t$ then if $\beta^{'2}\ge q_t$ --- hence with unconditional probability $t(1-t)$ --- $U_j$ is returned as uniform on $[t,1]$, and the square projection of $Y^{(j)}$ in the direction of $X$ is replaced by $\omega$ times $\beta^{'2}/n$. It keeps the same sign as $\beta$, and the orthogonal projection is rescaled to preserve the length of $Y$.
		\item With probability $(1-t)^2$ we have $\beta^2 \le q_t$ and $\beta^{'2}$ $U_j$ is returned as the beta cdf of $\beta^2/n$, and $Y$ is returned unchanged.
	\end{enumerate}
	In case 3 we have, trivially, $\mu_{j+1}=\mu$.
	In the other cases we have
	\begin{equation}
	    \label{E:Yj}
    Y^{(j+1)} = \sqrt{\omega} \frac{\beta'}{\sqrt{n}} X + \sqrt{\frac{n-\beta^{'2}}{n-\beta^2}}\left(Y^{(j)} - \sqrt{\omega}\frac{\beta}{\sqrt{n}} X \right).    
	\end{equation}
	Cases 1 and 2 produce different distributions of $\beta'$.
	They also differ in the distribution of $\beta$, $\bmu$, and $\bp$ implicitly, since the occurrence of case 1 depends on $\beta^2>q_t$.
	Let $M$ be the number of occurrences of either of these cases, and let $\eu{A}$ be the event that
	$$
	\{\beta_j^2,\beta_j^{'2} \le \sqrt{n} \text{ for all null } 1\le j\le p\} \cap \{M \le \sqrt{n} \}.
	$$
	Using equation \eqref{E:betaone} of Lemma \ref{L:betasum} for the first component and Bernstein's inequality for the second (as $M$ is binomial distributed with parameter $(p-k,2t-t^2)$, we have
	\begin{equation}
		\label{E:PB}
		\P(\eu{A}) \ge 1-2p\exp\left(-\frac38\sqrt{n}\right) - \exp\left(-\frac38 (\sqrt{n}-2tp)\right)
	\end{equation}
	In what follows we will condition on the event $\eu{A}$, which reduces the power by at most $1-\P(\eu{A})$.
	
	Projecting the equation \eqref{E:Yj} onto the direction $\hat\mu$ and dividing by $\sqrt{\omega}$ we get
	\begin{align*}
		\mu_{j+1} &= \frac{\beta' \beta^\mu}{n} + \sqrt{1+ \frac{\beta^2 -\beta^{'2}}{n-\beta^2}} \left(\mu_j - \frac{\beta\beta^\mu}{n} \right)\\
		& \ge \sqrt{1+ \frac{\beta^2 -\beta^{'2}}{n-\beta^2}} \left(1- \frac{\beta^{\mu 2}}{n} \right)\mu_j + \frac{\beta' \beta^\mu}{n} -
		\sqrt{1+ \frac{\beta^2 -\beta^{'2}}{n-\beta^2}} \frac{\bp \bmu}{n-1}\\
		&\ge \left( 1 - \frac{\beta^{'2}}{n} -\frac{\beta^{\mu 2}}{n} \right)\mu_j
		- \frac12 \frac{\beta^{'2} }{n}  - \frac34 \frac{\beta^{\perp 2} }{n}     - \frac34 \frac{\beta^{\mu 2} }{n},
	\end{align*}
	conditioned on $\eu{A}$, for $n>10$.
	Here we are using the fact that $\beta'$ and $\beta$ have the same sign, which implies that either $\beta'\beta^\mu$ or $-\beta^\perp \beta^\mu$ must be positive, so that one of these may be dropped from this lower bound; and the general inequality $4|xy|\le (x+y)^2$.
	
	Let $\eu{I}_j$ be the indicator of $X_j$ being a null predictor and case 1 or 2 holding for distillation step $j$; and $\eu{I}'_j$ be the indicator of $X_j$ being an active predictor.
	Define
	$$
	B^*:= \max \Bigl\{ \sum_{j=1}^p \eu{I}_j \beta_j^2, \sum_{j=1}^p \eu{I}_j \beta^{\mu 2}_j, \sum_{j=1}^p \eu{I}_j \beta_j^{\perp 2}, \sum_{j=1}^p (\eu{I}_j + \eu{I}'_j) \beta^{'2}_j \Bigr\}.
	$$
	Then for any $m,m'$ such that all predictors $X_j$ in $m<j\le m'$ are null,
	$$
	\mu_{m'} \ge \left( 1- \frac{2B^*}{n}\right) \mu_m -  \frac{2 B^*}{n}.
	$$
	
	Now consider what happens when the process hits a true predictor $X_j$.
	As regards the shrinking of $\mu_j$, the only change is that the $\mu$ component $b\sqrt{n}$ replaces $\beta^\mu$.
	Using the assumption on $n$, and the condition that $\eu{A}$ imposes on $\beta'$ we get
	\begin{align*}
		\mu_{j+1} &\ge \left(1-b^2- \frac{\beta^{'2}}{ n} \right) \mu_j - \frac{b|\beta'|}{\sqrt{n}} \mathbf{1}\{\beta'\beta^\mu<0\} -  \frac{b|\beta^{\perp}|}{\sqrt{n-1}} \mathbf{1}\{\beta^\perp\beta^\mu>0\}.
	\end{align*}
	Thus, if $j_l$ is the $l$-th true predictor, the size of the remaining signal when we test $X_{j_l}$ is at least
	\begin{equation} \label{E:mubound}
		\mu_{j_l} \ge \left( 1- \frac{2B^*}{n} \right) (1-b^2)^{l-1} \mu_1 -  \frac{2 B^*}{n} - \frac{b }{\sqrt{n-1}} \sum_{i=1}^l \max\{|\beta^{'}_{j_i}| , |\beta^{\perp}_{j_i}|\}.
	\end{equation}
	Note that the extra $\beta^{'2}$ terms corresponding to the true predictors have been absorbed into $B^*$, which is why the definition of $B^*$ included the extra indicator $\eu{I}'_j$ for the $\beta^{'2}$ sum.
	
	The p-value produced when testing true predictor $X_{j_l}$ is the same as $F_\beta(W_{j_l}^2/\omega)$, where $F_\beta$ is the complementary cdf of the Beta$(\frac12,\frac{n-1}{2})$ distribution.
	By Lemma \ref{L:betasum} $\log F_\beta(x) < \frac{9}{32} - \frac38\cdot n x$ for $x>1$, we have
	\begin{equation}
	    \label{E:xibound}
        	\xi \ge \sum_{l=1}^{k_*} \Bigl( \frac38 \frac{n}{\omega}W_{j_l}^2 - \frac{9}{32} \Bigr)
	\end{equation}
    if the summands are all bigger than $-\log t$.
    (This condition is required to avoid the possibility that one or more true predictors will yield a p-value that is above the threshold $t$, and thus will be swapped out for a new random uniform p-value.
    If $p$ is large the penalty for this swap will be small, or even negative, since there will be spurious null-predictor p-values below $t$ to take its place, but an assumption of large $p$ to improve the power is contrary to the sense of the test.)
    Note that we only sum up to $k_*= \min\{k, K\}$.
    
	Define $B^+ :=  \sum_{i=1}^{k-1}\max\{|\beta^{'}_{j_i}| , |\beta^{\perp}_{j_i}|\}$
    and $B^{++} := \max_{1\le i\le k_*} |\beta^\perp_{j_i}|$,
    and let $\eu{A}'$ be the event
	$$
	\eu{A}' := \{B^* \le \frac{n \mu_1}{3} \delta \} \quad \cap \quad \{ B^+ \le \frac{\sqrt{n-1}\mu_1}{3b} \delta \} \quad \cap\quad 
    \{ B^{++} \le \frac{\sqrt{n} \mu_1 b}{\sqrt{1-b^2}} \left( (1-b^2)^{k-1} - \delta \right) \}.
	$$
	Then on the event $\eu{A}\cap \eu{A}'$ the bound \eqref{E:xibound} holds; for all $l\in\{1,\dots,k\}$, by \eqref{E:mubound},
	$$
	\mu_{j_l} \ge (1-\frac{\delta}{3}) (1-b^2)^{l-1} \mu_1 - \frac{2}{3} \delta \mu_1 \ge \left((1-b^2)^{l-1} - \delta\right) \mu_1,
	$$
	and
$$
		|W_{j_l}|=  \sqrt{\frac{\omega}{n}} \bigl| \sqrt{n} b\mu_{j_l} + \sqrt{1-b^2}  \beta^\perp_{j_l} \bigr|.
$$
	Then (since $b\sqrt{1-b^2}\le \frac12$)
	\begin{align*}
		\xi &\ge  \frac38 \mu_1^2 \sum_{l=0}^{k_*-1} b^2 n  \left((1-b^2)^{2l} - 2(1-b^2)^{l}  \delta \right) - \sqrt{n} \mu_1 B^+ \\
		&\ge 
		 \frac38 n \mu_1^2 \left(\frac{1- (1-b^2)^{2k_*}}{2-b^2} - 2\delta \right) - \frac{9k}{32} \\
		&= \xi_\alpha
	\end{align*}
	by the definition of $\delta$.
	Thus the probability of $\{\xi\ge \xi_\alpha\}$ is bounded below by $\P(\eu{A}\cap \eu{A}')$.
	The result then follows by combining \eqref{E:PB} with
	\begin{align*}
		\P&(\eu{A}^{'\complement} \, | \, \eu{A})  \le \P\left(B^*  > \frac{n \mu_1}{3} \delta \right) \quad + \quad
		\P \left( B^+ > \frac{\sqrt{n-1}\mu_1}{3} \delta \right) \\
        &\hspace*{5cm} + \quad
        \P\left( B^{++} > \frac{\sqrt{n} \mu_1 b}{\sqrt{1-b^2}} \left( (1-b^2)^{k_*-1} - \delta \right) \right) \\
		&\le 4 \exp\left(-\frac{n\mu_1 \delta}{8} + \sqrt{n}\log\frac{\sqrt{n}\mu_1\delta}{4t}  \right)  + 12^{k_*-1}\exp\left(  - \frac{(n-1)\mu_1^2 \delta^2}{24(k_*-1)}\right)\\
        & \qquad + 2k \exp\left( -\frac{3 n \mu_1^2}{8(1-b^2)} \left[ b^2 \left( (1-b^2)^{k-1} -\delta \right)^2 -
        2(n \mu_1^2)^{-1/2} (1-b^2)^{k_*-1}\sqrt{1 - 3 \log t } \right]
        \right).
	\end{align*}
	The first bound uses \eqref{E:betasum} of Lemma \ref{L:betasum} with $m\le 4tp$; the second uses \eqref{E:betasum2} with $m=k-1$,
	and the simplification $(n-1)/24 \ge n/26$ for $n\ge 13$;
    and the third uses \eqref{E:betaone}.
\end{proof}

\begin{lemma}
	\label{L:betasum}
	Let $B_1,\dots,B_m$ be an i.i.d.\ sequence of random variables such that $B_j$ has $\operatorname{Beta}(\frac12,\frac{n-1}{2})$ distribution.
	Let $q_t$ be the probability $t$ quantile of this distribution.
	Then for any $x>0$,
	\begin{equation}
		\label{E:betaone}
		\log \P\bigl( n B_j > x \bigr) \le \frac{21}{32} - \frac38 x
	\end{equation}
	and for $x>mq_t$
	\begin{equation}
		\label{E:betasum}
		\log \P\bigl( \sum_{j=1}^m nB_j > x  \, \bigm| \, \text{all } B_j > q_t \bigr) \le
		  -\frac 38 x  + m \log \frac{2x}{mt} .
	\end{equation}
	Finally, letting $B'_1,\dots,B'_m$ be an independent sequence with distribution $\operatorname{Beta}(\frac12,\frac{n'-1}{2})$,
	\begin{equation}
		\label{E:betasum2}
		\log \P\bigl( \sum_{j=1}^m \max\{\sqrt{nB_j}, \sqrt{n'B'_j} \} >x \bigr) \le
		 - \frac38  \frac{x^2}{m}  + m \log 12.
	\end{equation}
\end{lemma}

\begin{proof}
	\eqref{E:betaone} is a direct consequence of formula (4) of \cite{mS23}.
	The MGF of $B_j$ is then bounded above by
	$$
	  M(\lambda) \le \ee^{21/32} \frac{3}{3-8\lambda}
	$$
	for $\lambda<\frac38$.
	Using the independence and Chernoff's inequality we then have
	$$
		\log \P\bigl( \sum_{j=1}^m B_j > x \bigr) \le \frac{21}{32} m - m\log\left(1- \frac{8}{3} \lambda \right)-\lambda x.
	$$
	This is minimized at $\lambda = \frac38-\frac{m}{x}$, yielding
	$$
		m^{-1}\log \P\bigl( m^{-1}\sum_{j=1}^m B_j > x \bigr) \le \frac{21}{32}m- m \log\left(\frac{8}{3x} \right)-\frac38 x + m\le m \log \frac{2x}{m} - \frac38x.
	$$	
	The result \eqref{E:betasum} then follows since $\P(B_j > q_t)=t$.
	
	The final bound is very similar, but now the random variable $\max\{\sqrt{B_j}, \sqrt{B'_j} \}$ is subgaussian with MGF bounded by
	$$
		M(\lambda) \le 2\ee^{21/32} \sqrt{8\pi/3} \ee^{2\lambda^2/3}\le 12 \ee^{2\lambda^2/3}.
	$$
	Applying Chernoff's inequality now with $\lambda=\frac{3x}{4m}$ yields \eqref{E:betasum2}.
\end{proof}

\subsection{Proof of Lemma \ref{lem:stick_breaking}}
{\ }\\
{\em Statement of Lemma}:
For an $n$-dimensional random vector $W$ let $\omega := \left|\left|W\right|\right|^2_2$ and $B_j := W_j^2 \ / \ \sum\limits_{i = j}^n W_j^2$ for $j = 1,\ldots,n-1$. Let $\mu : \mathbb{R}^n \rightarrow \mathcal{G} $ be the bijective map \begin{equation}
	\mu(W) = \left(B_1,\ldots B_{n-1}, \sgn\left(W_1\right),\ldots,\sgn\left(W_n\right), \omega \right) = \left(B, \sgn\left(W\right), \omega\right).
\end{equation} If $W \sim N(0,I)$, then all elements of $\mu(W)$ are mutually independent where $\sgn\left(W_j\right) \sim \mathrm{Rademacher}$, $B_j \sim \mathrm{Beta}\left(\frac{1}{2},\frac{n-j}{2}\right)$, and $\omega \sim \chi^2_{n}$.

{\ }\\
\noindent {\em Proof:} \quad
Let $S_j = \sum\limits_{i = j}^{n} W_i^2$, so $B_j = W_j^2 / S_j$ and $S_1=\omega$. 
By the symmetry of the Gaussian distribution, the collections of random variables $\{|W_j|: \, j=1,\dots,n\}$ and $\{\sgn(W_j): \, j=1,\dots,n\}$ are independent of each other.
Hence we need only show that $ \{ B_1,\ldots B_{n-1},\omega\}$ are mutually independent.

We begin by noting the following facts:
\begin{enumerate}
	\item Lukacs's proportion-sum Independence Theorem: $B_j$ is independent of $S_j$, and $B_j \sim \mathrm{Beta}\left(\frac{1}{2},\frac{n-j}{2}\right)$.
	\item For each $j\ge 2$, $B_j\indep (B_1,\dots, B_{j-1},S_1,\dots,S_{j-1})$ conditioned on $S_j$.
	This follows from the fact that the $W_i$'s are mutually independent and $S_1,\dots,S_{j-1}, B_1,\dots,B_{j-1}$
	are all measurable with respect to $(S_j, W_1,\dots, W_{j-1})$.
\end{enumerate}

We now prove by induction on $j$ that for any $j\in \{0,\dots,n-1\}$ the random variables $\omega, B_1,\dots,B_j $ are mutually independent.
For $j=0$ the statement is trivial.
Assume now that $\omega,B_{1},\dots,B_{j}$ are mutually independent.
Consider any bounded measurable functions $f_0,f_1,\dots,f_j$.
Then

\begin{align*}
	\E\left[f_0(\omega) \prod_{i=1}^{j} f_i (B_i) \right] &=\E\left[ \E\left[f_0(S_1) \prod_{i=1}^j f_i (B_{i}) \, \bigm| \, S_j \right] \right]\\
	& = \E\left[ \E\left[f_0(S_1) \prod_{i=1}^{j-1} f_i (B_{i}) \, \bigm| \, S_j \right] \cdot \E\left[ f_j (B_{j}) \, \bigm| \, S_j \right] \right] \text{ by fact 2}\\
	& = \E\left[ \E\left[f_0(S_1) \prod_{i=1}^{j-1} f_i (B_{i}) \, \bigm| \, S_j \right] \cdot \E\left[ f_j (B_{j})  \right] \right] \text{ by fact 1} \\
	& =  \E\left[f_0(S_1) \prod_{i=1}^{j-1} f_i (B_{i})\right] \cdot \E\left[ f_j (B_{j})  \right]\text{ by the law of total expectation} \\
	&=   \E\left[f_0(S_1)\right] \cdot \prod_{i=1}^{j-1} \E \left[ f_i (B_{i})\right] \cdot \E\left[ f_j (B_{j})  \right]\text{ by the induction hypothesis} . \qed
\end{align*}

\clearpage
\section{Power Curves} \label{sec:power_curves}

\begin{figure}[H]
	\centering
	\includegraphics[width=0.98\linewidth]{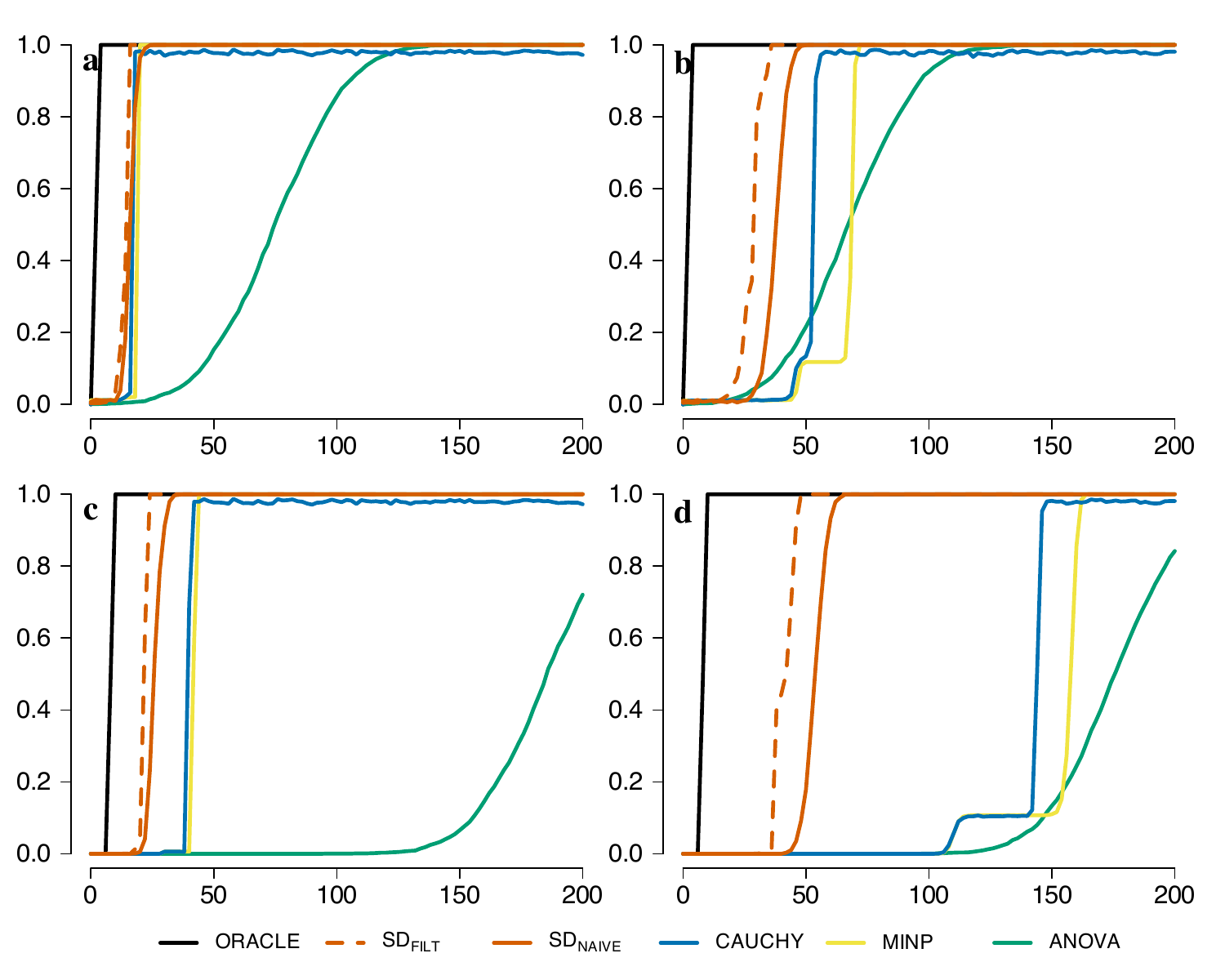}
	\caption{\textbf{Power in the dense $X$, $n>p$ case where $r^2 = 0.2$.} Power plotted as a function of target signal strength $s$. $n = 10^5$ and $p = 10^4$. (\textbf{a}) 4 active predictors with type-1 error rate $10^{-2}$. (\textbf{b}) 16 active predictors with type-1 error rate $10^{-2}$. (\textbf{c}) 4 active predictors with type-1 error rate $10^{-8}$.  (\textbf{d}) 16 active predictors with type-1 error rate $10^{-8}$.}
	\label{dense_large_n_2}
\end{figure}

\begin{figure}[H]
	\centering
	\includegraphics[width=0.98\linewidth]{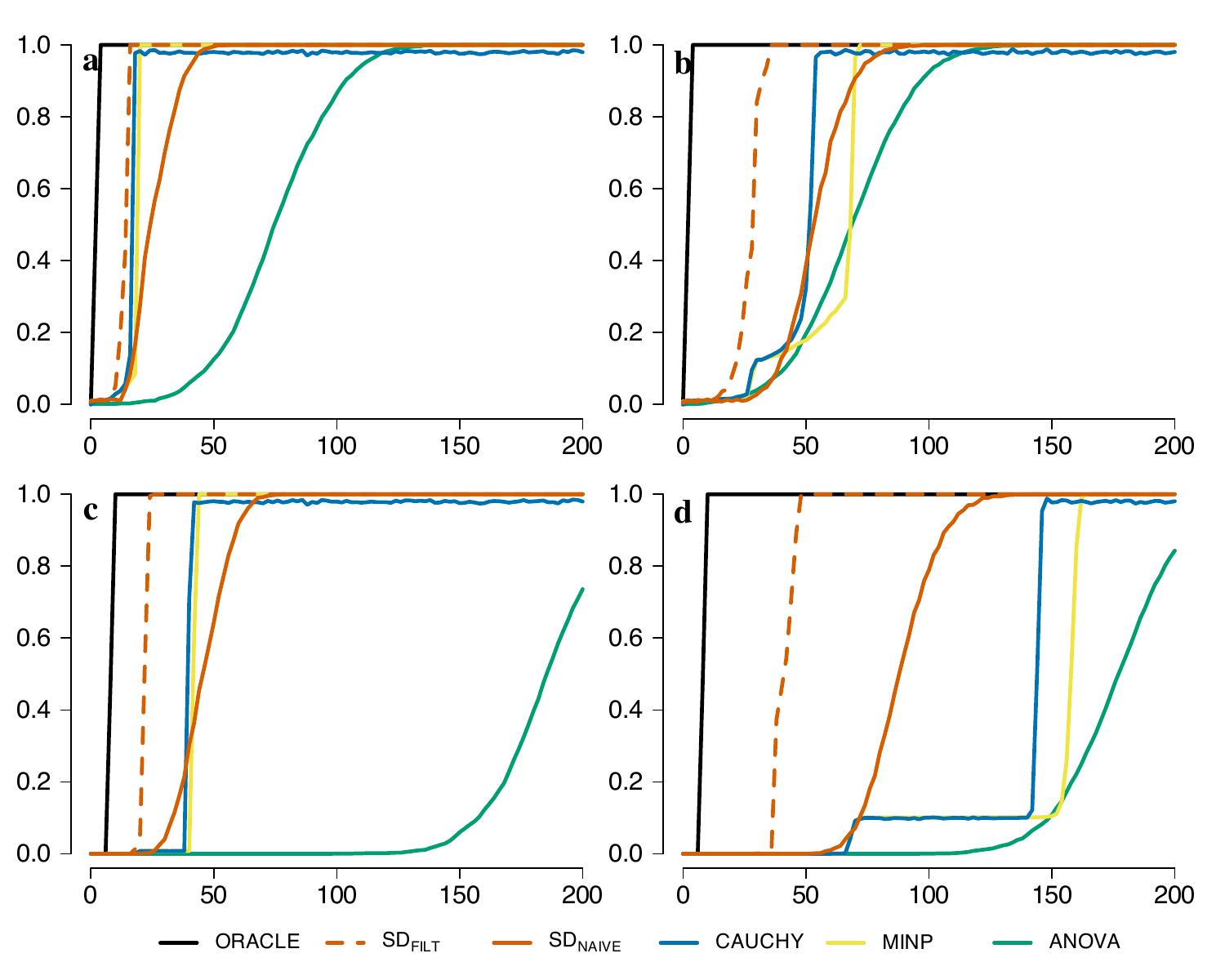}
	\caption{\textbf{Power in the dense $X$, $n>p$ case where $r^2 = 0.5$.} Power plotted as a function of target signal strength $s$. $n = 10^5$ and $p = 10^4$. (\textbf{a}) 4 active predictors with type-1 error rate $10^{-2}$. (\textbf{b}) 16 active predictors with type-1 error rate $10^{-2}$. (\textbf{c}) 4 active predictors with type-1 error rate $10^{-8}$.  (\textbf{d}) 16 active predictors with type-1 error rate $10^{-8}$.}
	\label{dense_large_n_5}
\end{figure}

\begin{figure}[H]
	\centering
	\includegraphics[width=0.98\linewidth]{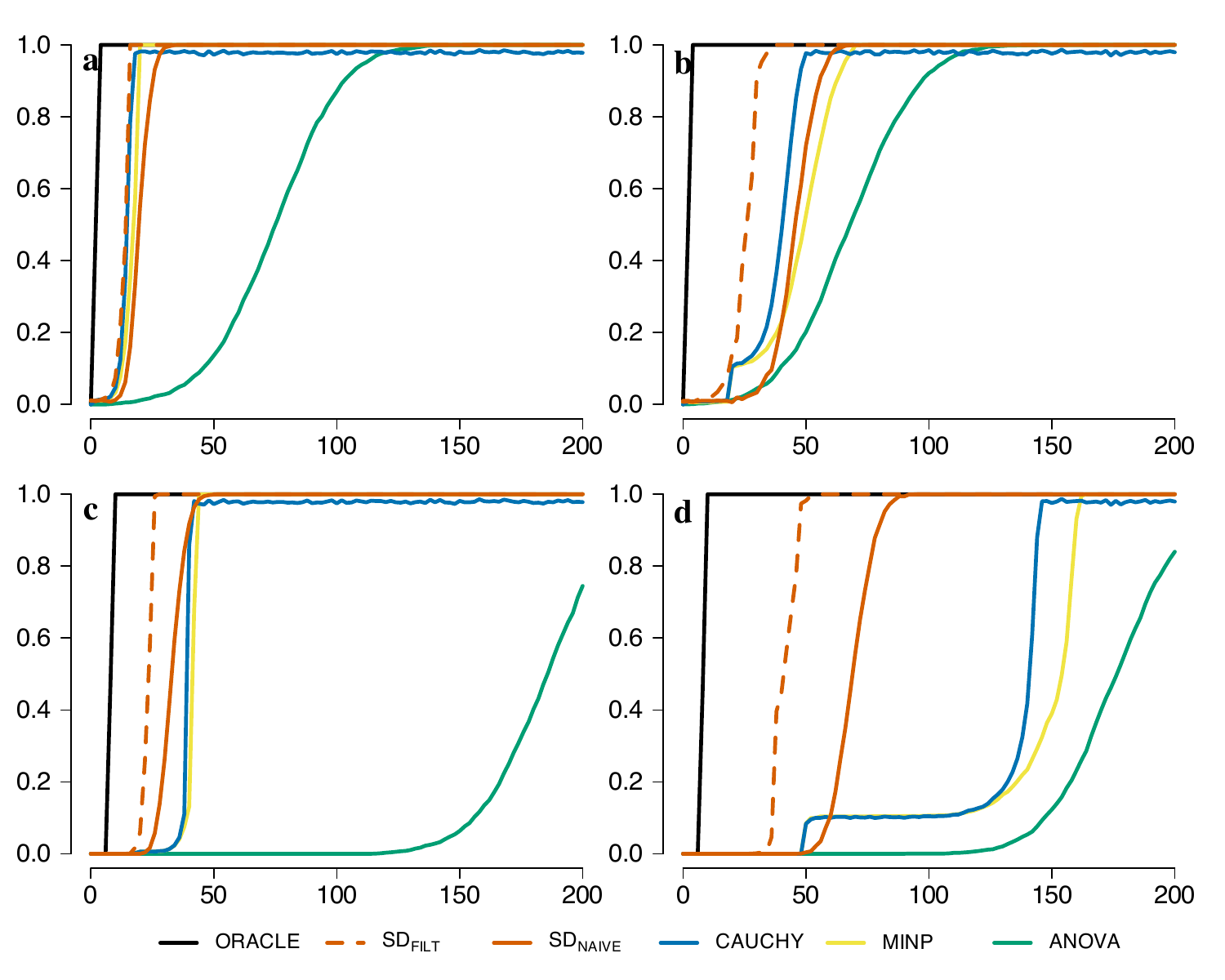}
	\caption{\textbf{Power in the dense $X$, $n>p$ case where $r^2 = 0.8$.} Power plotted as a function of target signal strength $s$. $n = 10^5$ and $p = 10^4$. (\textbf{a}) 4 active predictors with type-1 error rate $10^{-2}$. (\textbf{b}) 16 active predictors with type-1 error rate $10^{-2}$. (\textbf{c}) 4 active predictors with type-1 error rate $10^{-8}$.  (\textbf{d}) 16 active predictors with type-1 error rate $10^{-8}$.}
	\label{dense_large_n_8}
\end{figure}

\begin{figure}[H]
	\centering
	\includegraphics[width=0.98\linewidth]{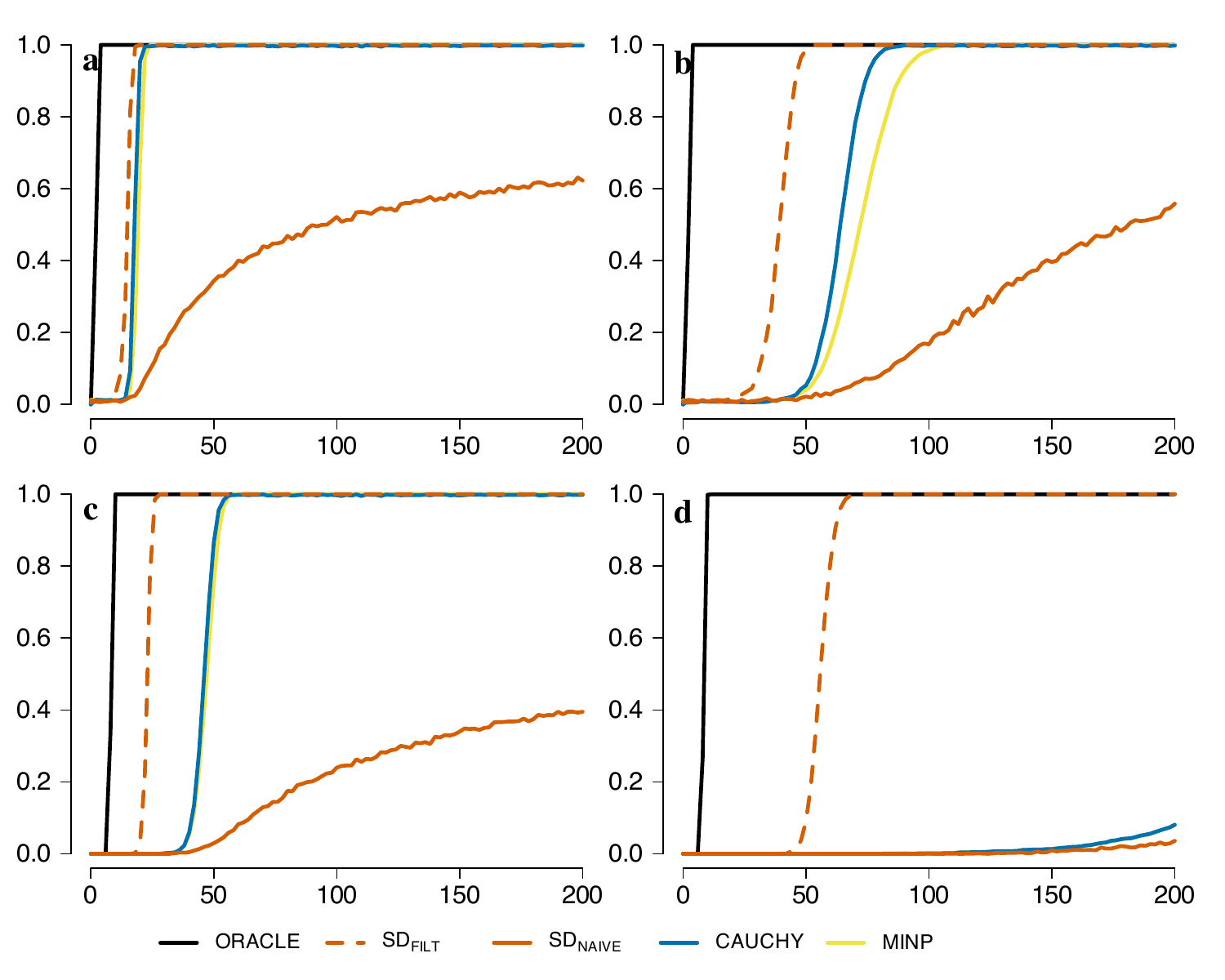}
	\caption{\textbf{Power in the dense $X$, $n<p$ case where $r^2 = 0.2$.} Power plotted as a function of target signal strength $s$. $n = 10^3$ and $p = 10^4$. (\textbf{a}) 4 active predictors with type-1 error rate $10^{-2}$. (\textbf{b}) 16 active predictors with type-1 error rate $10^{-2}$. (\textbf{c}) 4 active predictors with type-1 error rate $10^{-8}$.  (\textbf{d}) 16 active predictors with type-1 error rate $10^{-8}$.}
	\label{dense_small_n_2}
\end{figure}

\begin{figure}[H]
	\centering
	\includegraphics[width=0.98\linewidth]{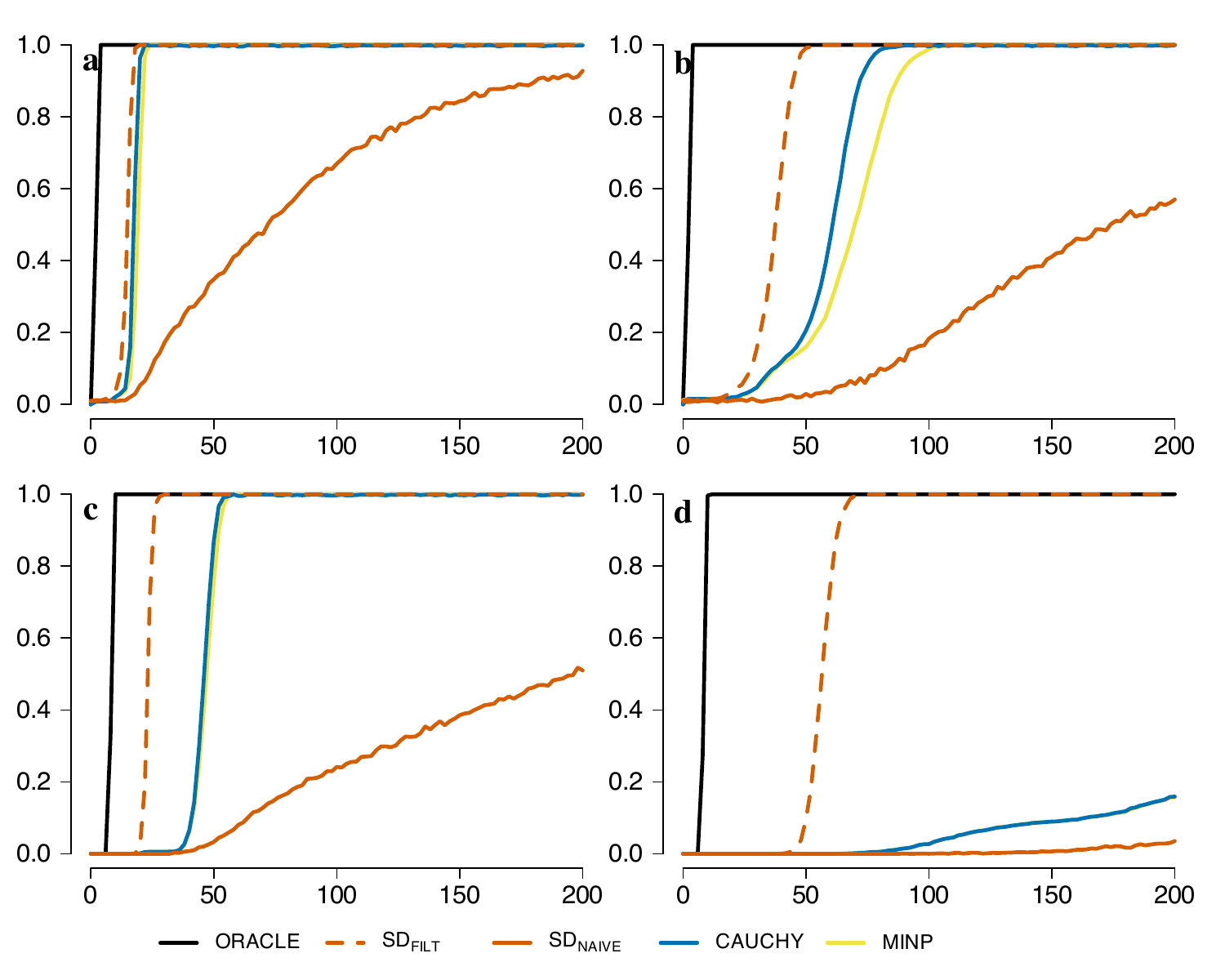}
	\caption{\textbf{Power in the dense $X$, $n<p$ case where $r^2 = 0.5$.} Power plotted as a function of target signal strength $s$. $n = 10^3$ and $p = 10^4$. (\textbf{a}) 4 active predictors with type-1 error rate $10^{-2}$. (\textbf{b}) 16 active predictors with type-1 error rate $10^{-2}$. (\textbf{c}) 4 active predictors with type-1 error rate $10^{-8}$.  (\textbf{d}) 16 active predictors with type-1 error rate $10^{-8}$.}
	\label{dense_small_n_5}
\end{figure}

\begin{figure}[H]
	\centering
	\includegraphics[width=0.98\linewidth]{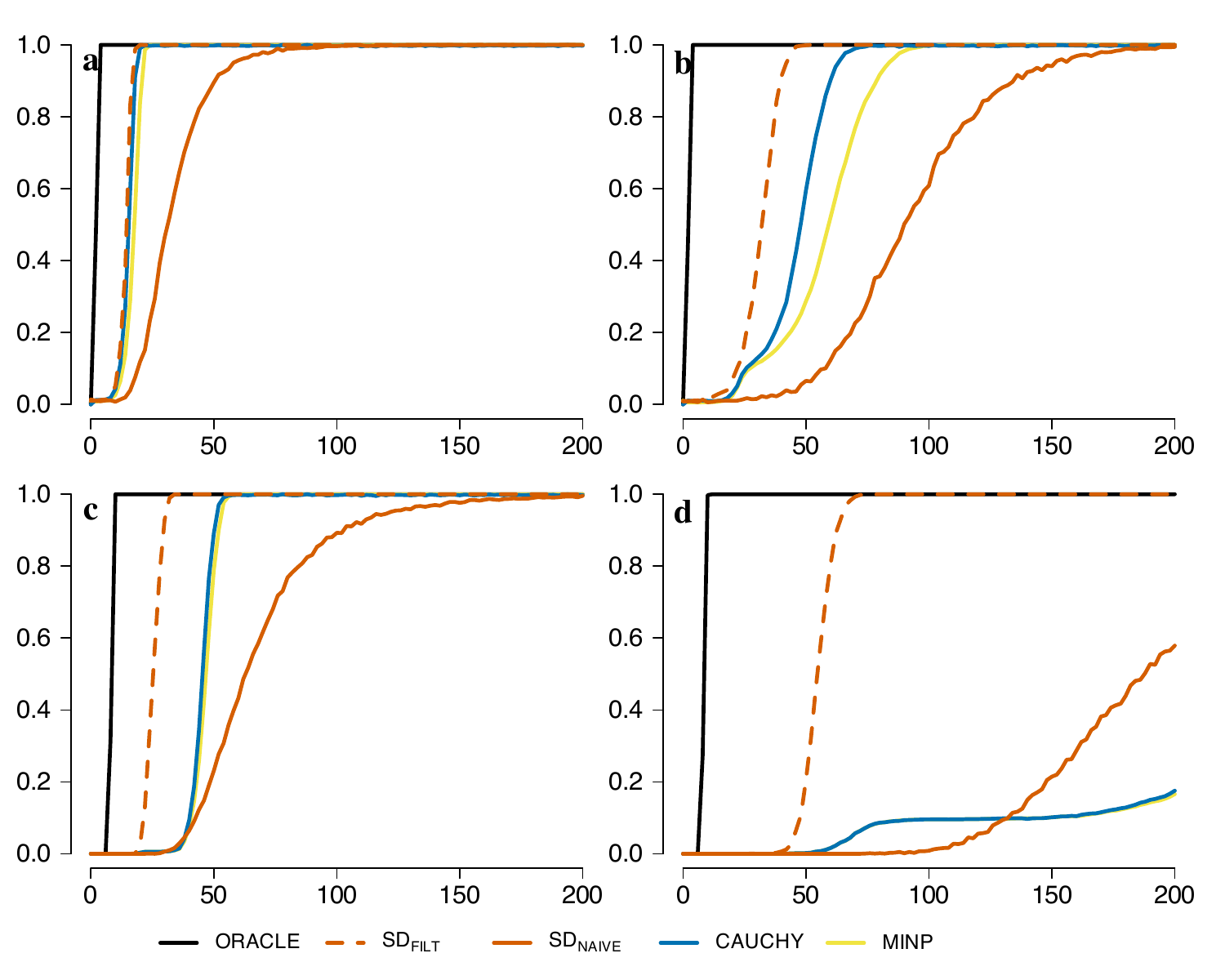}
	\caption{\textbf{Power in the dense $X$, $n<p$ case where $r^2 = 0.8$.} Power plotted as a function of target signal strength $s$. $n = 10^3$ and $p = 10^4$. (\textbf{a}) 4 active predictors with type-1 error rate $10^{-2}$. (\textbf{b}) 16 active predictors with type-1 error rate $10^{-2}$. (\textbf{c}) 4 active predictors with type-1 error rate $10^{-8}$.  (\textbf{d}) 16 active predictors with type-1 error rate $10^{-8}$.}
	\label{dense_small_n_8}
\end{figure}
\newpage

\end{document}